\documentclass[11pt]{article}
\usepackage[utf8]{inputenc}
\usepackage{amssymb}
\usepackage{amsmath}
\usepackage{graphicx}
\usepackage{verbatim}
\usepackage{comment}
\usepackage[margin=0.75in]{geometry}
\usepackage{caption}
\usepackage{subcaption}
\usepackage{placeins}
\usepackage{bookmark}
\usepackage{hyperref}
\usepackage{graphicx}
\usepackage{float}
\usepackage{amsfonts}
\usepackage{amsmath,amsthm,amssymb}
\usepackage{amsmath,amssymb}
\usepackage{epstopdf}
\usepackage{algorithm, algorithmic}
\usepackage{color}
\usepackage{wrapfig}
\usepackage{marvosym}
\usepackage{caption}
\usepackage{cleveref}
\usepackage{ragged2e}
\usepackage[scaled]{helvet} 
\usepackage{fourier}
\usepackage{mathtools}
\usepackage{verbatim}

\newcommand{\bF}{{\boldsymbol{F}}}

\newcommand{\bx}{\boldsymbol{x}}
\newcommand{\bPsi}{\boldsymbol{\Psi}}

\newcommand{\bA}{\boldsymbol{A}}
\newcommand{\by}{\boldsymbol{y}}

\newcommand{\bS}{\boldsymbol{S}}

\newcommand{\be}{\boldsymbol{e}}

\newcommand{\ba}{\boldsymbol{a}}

\newcommand{\bg}{\boldsymbol{g}}

\newcommand{\bz}{\boldsymbol{z}}

\newcommand{\bOmega}{\boldsymbol{\Omega}}

\newcommand{\samp}{\mathrm{samp}}

\newcommand{\Fu}{\boldsymbol{F_u}}

\DeclarePairedDelimiter{\normOne}{\lVert}{\rVert_1}
\DeclarePairedDelimiter{\normTwo}{\lVert}{\rVert_2}
\newcommand{\bW}{\boldsymbol{W}}

\newcommand{\bX}{\boldsymbol{X}}
\newcommand{\bY}{\boldsymbol{Y}}

\title{A Practical Study of Longitudinal Reference Based Compressed Sensing for MRI}
\date{\today}
\author{Samuel Birns, Bohyun Kim, Stephanie Ku, Kevin Stangl, Deanna Needell}

\begin{document}
\maketitle

\begin{abstract}
Compressed sensing (CS) is a new signal acquisition paradigm that enables the reconstruction of signals and images from a low number of samples.  A particularly exciting application of CS is Magnetic Resonance Imaging (MRI), where CS significantly speeds up scan time by requiring far fewer measurements than standard MRI techniques.  Such a reduction in sampling time leads to less power consumption, less need for patient sedation, and more accurate images. This accuracy increase is  especially pronounced in pediatric MRI where patients have trouble being still for long scan periods.  Although such gains are already significant, even further improvements can be made by utilizing past MRI scans of the same patient. Many patients require repeated scans over a period of time in order to track illnesses and the prior scans can be used as \textit{references} for the current image. This allows samples to be taken \textit{adaptively}, based on both the prior scan and the current  measurements.
 Work by Weizman \cite{weizman2015compressed} 
has shown that so-called \textit{reference based adaptive-weighted temporal Compressed Sensing MRI} (LACS-MRI) requires far fewer samples than standard Compressed Sensing (CS) to achieve the same reconstruction signal-to-noise ratio (RSNR). The method uses a mixture of reference-based and adaptive-sampling.  In this work, we test this methodology by using various adaptive sensing schemes, reconstruction methods, and image types. We create a thorough catalog of reconstruction behavior and success rates that is interesting from a mathematical point of view and is useful for practitioners. We also solve a grayscale compensation toy problem that supports the insensitivity of LACS-MRI to changes in MRI acquisition parameters and thus showcases the reliability of LACS-MRI in possible clinical situations.  
 
 \end{abstract}

\section{Introduction}

\indent  Magnetic Resonance Imaging (MRI) has become a staple of modern medicine. Because it is noninvasive and it produces high quality images, MRI is used widely in both treatment and research ~\cite{hollingworth2000diagnostic}. However, a typical MRI can take upwards of an hour and requires the patient to be completely still; even the slightest movement will induce enough noise in the image to make the reconstruction unusable~\cite{kirschen1989use}. For this reason, advancements that decrease the amount of time a MRI image takes are incredibly beneficial.  MRI is an ideal application of Compressed Sensing (CS), a method that enables signal reconstruction with only a fraction of the original information by numerically solving under-determined linear systems ~\cite{RefWorks:309}. Since we only need to collect a small percentage of the original data, CS can make MRI much faster~\cite{RefWorks:70}. This is possible because MRI images can be represented in a $\textit{sparse}$ way; that is, there are linear maps that transform these images into matrices with only a small number of non-zero entries. 

\subsection{Mathematical Formulation}
 CS relies on solving on under-determined linear systems using mathematical optimization.  Consider a signal (e.g. vector, matrix or image) $\bx\in\mathbb{C}^n$ that has $s \ll n$ non-zero entries.  We call such a signal $s$-sparse, and write $\|\bx\|_0 := |\{i : \bx_i\ne 0\}| = s << n$.  We take (noisy) linear measurements of $\bx$ of the form $\by = \bA\bx + \be$, where $\bA$ is an $m\times n$ measurement matrix, $\be\in\mathbb{C}^m$ is a gaussian noise term, and $\by\in\mathbb{C}^m$ is called the measurement vector.  When $m\ll n$, $\by$ can be used to efficiently recover most of the information in $\bx$.  The goal of CS is to (i) design efficient measurement matrix types $\bA$ that capture the most important information in desired signals, and (ii) design reconstruction methods that recover the signal $\bx$ from its measurements $\by$.  Since the recovery problem is highly underdetermined, identifying the true signal $\bx$ is non-trivial, and only becomes a well-posed problem under the assumption of sparsity.  For example, whenever $\bA$ is one-to-one on sparse signals and there is no noise in the measurements (i.e. $\be=0$), one can solve the $\ell_0$-minimization problem to accurately recover the signal $\bx$:
\begin{equation}
	 \underset{\bg \in \mathbb{C}^N}{\text{min}} \, \left \Vert \bg \right \Vert_{0} \text{ s.t. } \, \,           \by = \bA\bg. \\
	\label{Candes L_0 optimization equation}
\end{equation}
Unfortunately in general, this problem is NP-Hard~\cite{muthukrishnan2005data}, so more practically useful recovery methods are needed.

Cand\`es et. al. ~\cite{emmanuel2004robust} have developed two fundamental results that link sparsity to theoretical image reconstruction.
 The authors introduced the Restricted Isometry Property (RIP) ~\cite{berman1988matrix}. A matrix is said to satisfy the RIP if it is nearly isometric (i.e. distance preserving) on sparse vectors; specifically, an $m \times n$ matrix $\bA$ satisfies the RIP (of order $s$) if there exists $\delta_s \in (0,1)$ such that for all $m \times s$ submatrices $\bA_s$ of $\bA$ and for every $s$-sparse vector $\bx$,
\begin{align}
	(1 - \delta_s) \Vert \bx \Vert^2_{2} \le \left \Vert \bA_s \bx \right \Vert^2_{2} \le (1 + \delta_s)               \Vert \bx \Vert^2_{2}
	\label{RIP}
\end{align}

An early result in CS \cite{candes2005error,RefWorks:285} states that, given a set of randomly selected indices $\bOmega$ from $\{1, 2, \ldots, n\}$ with $|\bOmega| \approx s \log^4{N}$, if $\bF_u$ is the subsampled Fourier transform (i.e. the $m\times n$ submatrix of the discrete Fourier transform (DFT) whose rows correspond to the indices in $\bOmega$), then with high probability $\bF_u$ satisfies the RIP, and the optimization problems
\begin{equation}
	 \underset{\bg \in \mathbb{C}^N}{\text{min}} \, \left \Vert \bg \right \Vert_{0} \text{s.t.} \, \,           \bg|_{\bOmega} = \bF_u \quad\text{and}\quad
	 \underset{\bg \in \mathbb{C}^N}{\text{min}} \, \left \Vert \bg \right \Vert_{1} \text{s.t.} \, \,           \bg|_{\bOmega} = \bF_u
	\label{Candes L_1 optimization equation}
\end{equation} are equivalent in the sense that these optimization problems yield the same solutions. Since $\ell_1$ optimization can be cast as an efficient linear programming problem, the ability to solve $\ell_1$ optimization in lieu of $\ell_0$ optimization has made CS much more practically useful. In addition, Cand\`es et. al. ~\cite{candes2006stable} show that if a matrix $A$ satisfies the RIP and if a vector $\bx_0 \in \mathbb{R}^m$ is $s$-sparse, given noisy measurements $\by = \bA \bx_0 + \be$, the solution $\hat{\bx}$ to the optimization problem 

\begin{align}
	\underset{\bx} {\text{min}} \left \Vert \bx \right \Vert_{1} \text{s.t.} \, \, \left \Vert \bA \bx -          \by \right \Vert_{2} \le \epsilon,
	\label{L1 minimization}
\end{align}
satisfies
\begin{align}
	\left \Vert \hat{\bx} - \bx_0 \right \Vert_{2} \le C \cdot \epsilon,
	\label{L1 minimization reconstruction bound}
\end{align}

where $\epsilon \geq \Vert e \Vert_{2}$.  
The combination of the two previous equations gives CS its utility since $\ell_1$-minimization produces  accurate signal recovery from highly underdetermined measurements.

Closely related to the RIP is the $\textit{coherence}$ of a matrix. Intuitively, the coherence of a matrix is a measure of how similar its columns are; the higher the coherence, the more similar its columns. Formally, coherence of a matrix $\bA$ is defined as 
\begin{align}
	\mu(\bA) = \underset{j < k}{\text{max}} \frac{| \langle \bA_j, \bA_k \rangle |}{ \left \Vert \bA_j \right          \Vert_2 \left \Vert \bA_k \right \Vert_2}
	\label{coherence}
\end{align} 
where $\langle \cdot \rangle$ denotes the standard inner product, and we write $\bA_j$ to denote the $j$th column of $\bA$. %\cite compressed sensing with coherent and redundant dictionaries 
Note that $0 \le \mu(\bA) \le 1$ for any matrix $\bA$ and that the closer $\mu(\bA)$ is to 0 (or the more $\textit{incoherent}$ $\bA$ is), the closer the columns of $\bA$ are to being orthogonal. Thus, coherence - or, more accurately, incoherence - and the RIP are very closely related. 

\subsection{Sparse representations}
The above theory and formulation hold for signals which are sparse.  However, most signals in practice are not themselves sparse, but rather can be represented sparsely in some basis.  For example, natural images are sparse in wavelet bases~\cite{Daube_Ten}; that is, we can represent an image $\bx$ by $\bx = \bW\bz$ for some sparse (or nearly sparse) signal $\bz$ and an orthonormal matrix $\bW$.  Since for any Gaussian matrix $\bA$ and orthonormal basis $\bW$, the product $\bA\bW$ is also Gaussian, the results above hold for Gaussian measurement matrices even when the signal is represented sparsely in an orthonormal basis.  For other types of RIP matrices, one requires that the RIP measurement matrix and the sparsifying orthonormal basis be incoherent \cite{RauhuSV_Compressed}.  In the MRI setting, the measurement matrix is a subsampled Fourier matrix and the sparsifying basis is typically a wavelet basis; unfortunately, these ensembles are not incoherent.
To overcome this problem and prove theoretical bounds on reconstruction while using wavelet bases, Krahmer and Ward~\cite{krahmer2014stable} use an approach known as \textit{variable density sampling} (VDS), which we describe in detail below.  This type of sampling stems from the idea that low frequency components contain the majority of the information about the underlying signal, so sampling primarily near the origin in $k$-space\footnote{We use the $k$-space representation of frequency space, where the low frequencies reside at the origin.  Subsampling a discrete Fourier ensemble is then equivalent to sampling coordinates $k_y$ from this $k$-space.} is more efficient and accurate than uniform random subsampling~\cite{lustig2007sparse}. Specifically, Krahmer and Ward~\cite{krahmer2014stable} define a novel probability density function that relates frequencies and $k$-space coordinates and is maximized at the origin. They then showed that the error from the reconstruction using this pdf is bounded.

\subsection{Longitudinal MRI}
While there is currently a wide variety of literature on CS and its uses in MRI, most of it does not take into account that often a patient who undergoes an MRI will undergo several MRIs spaced out over regular intervals of time. Since these time intervals are relatively short, i.e. often less than a month, it can be expected that there is little baseline change from one scan to the next. Weizman et. al. exploit this by introducing a reference image (a previous MRI scan) to classical CS reconstruction techniques~\cite{weizman2015compressed}. Specifically, Weizman et. al. iteratively solve:
\begin{align}
	\underset{\bx}{\text{min}} \, \normOne{\bW_1 \bPsi \bx} + \lambda \normOne{ \bW_2 (\bx - \bx_0)} \quad       
	\text { s.t. } \quad \normTwo{ \bF_u \bx - \by } < \epsilon,
	\label{lacs1}
\end{align} 
where $\bx_0$ is a reference image, $\by = \bF_u\bx_\star$ is the (possibly noisy) measurement vector for the current image $\bx_\star$, $\bPsi$ is a sparsifying operator such as the wavelet transform, $\bW_1$ and $\bW_2$ are diagonal weighted matrices (to be described momentarily), $\bF_u$ is the undersampled Fourier transform operator, and $\lambda$ is a parameter that can be modified to either emphasize sparsity (by making $ \Vert \bPsi \bx \Vert_1$ dominant in the minimization problem) or fidelity with the ground truth (by making $\Vert \bx - \bx_0 \Vert_1$ dominant). To solve the minimization problem, Weizman et. al. define an algorithm which they refer to as Longitudinal Adaptive Compressed Sensing MRI (LACS-MRI). 

\indent In each step of LACS-MRI, (\ref{lacs1}) is solved and new samples are selected using the discrete probability density function
\begin{align}
	f_S(k_y) = \gamma f_{ND}(k_y) + (1 - \gamma) f_{VD} (k_y).
	\label{mixedPDF}
\end{align} 
$f_{VD}$ is a variable density sampling probability density function that is maximized at $k_y = 0$.
 $f_{ND}$ is a density function designed to locate missing frequencies in the current signal estimation and $0 \le \gamma \le 1$ is a fidelity term that relates the reference and current image.  The weighting matrices $\bW_1$ and $\bW_2$ are updated based on the reconstruction in order to emphasize important signal entries, and $\gamma$ is updated based on the entries of $\bW_2$.
 
 \indent  By construction, when $\gamma \approx 0$, the $f_{VD}$ term dominates and a new set of $k$-space coordinates near the origin is sampled; conversely, when $\gamma \approx 1$, the $f_{ND}$ term dominates and the next iteration of $k$-space coordinates are taken from almost exactly the same region as the previous set of $k$-space coordinates. LACS-MRI updates $\gamma$ such that $\gamma \approx 0$ if the reconstruction is far from the ground truth and $\gamma \approx 1$ if the reconstruction is close to the ground truth.  

LACS-MRI is innovative since it uses both variable density sampling and adaptive sampling, where adaptive sampling is simply the process of choosing samples based off previous reconstructions, as opposed to sampling the same values regardless of previous measurements or iterations. While adaptive sampling may appear to be a logical way to drastically improve reconstruction, Davenport et. al. \cite{davenport2015constrained} shows that, perhaps counterintuitively, the performance of adaptive sampling is theoretically limited. In fact, they show that even in the best case, adaptive sampling reconstruction has an error term roughly  of $s \, \text{log}^2 \, n$, where $s$ is the sparsity of the target image and $n$ is the dimension of the target image, which is far lower than was previously conjectured. However, Davenport et. al. also show that, despite this theoretical result, when samples are forced to be chosen from a set of predetermined measurements, a property they refer to as \textit{constrained} measurements, adaptive sampling does indeed offer significant improvement over nonadaptive sampling.  

\indent Given an estimate of the signal's support, they provide an optimization problem that adaptively selects the optimal measurements from the Fourier ensemble (i.e. the optimal $k$-space frequencies to be used for the measurements).
This work is particularly relevant here as MRI measurements are inherently constrained, since measurements must be taken in $k$-space.  In this paper, we explore this type of adaptive sampling within the LACS-MRI framework, and compare various sampling techniques and reconstruction approaches.

\subsection{Contributions}  
We expand upon the work above by investigating the effects of combining variable density sampling and adaptive sampling into one algorithm. We study this algorithm  empirically by testing its performance under various sampling schemes. Additionally, we implement several modifications motivated by practical applications, including an approach that addresses different grayscale calibrations across MRI scans.

\section{LACS-MRI}
In this section, necessary background and details will be developed about LACS-MRI. LACS-MRI uses past reference images to help reconstruct the follow-up images for patients who require multiple brain scans. If the previous and current scans are similar, we may accelerate the recovery of follow-up scan and improve the accuracy ~\cite{weizman2014application}. Sometimes the reference image is not similar to the follow-up image due to surgery or treatment.  In most cases the underlying structure of the brain does not change very much so we can still benefit from the reference image.  Recall that the algorithm provides a reconstruction by solving the convex optimization problem:

\begin{equation} 
\hat{\bx} = \underset{\bx}{\text{argmin}} \, \underbrace{\normOne{\bW_1 \bPsi \bx}}_{\text{term 1}} + \underbrace{\lambda \normOne{ \bW_2 (\bx - \bx_0)}}_{\text{term 2}} \,       
	\text { s.t. } \, \, \, \normTwo{ \bF_u \bx - \by } < \epsilon,
	 \label{optproblem}
\end{equation}
where $\bx_0$ is the reference image,  $\hat{\bx}$ is our estimation of the  follow-up brain image, and $\by$ is the vector of complex measurements in $k$-space of the true follow-up brain image $\bx_\star$. $\bPsi$ is a sparsifying wavelet transform matrix and the $\Fu$ term is the sampling operator, which in our case represents line samples of $k$-space in 2D imaging. $\bW_1$ and $\bW_2$ are diagonal weight matrices and their $i$th entries, $[\cdot]_i$, are calculated as follows:

\begin{equation} 
\label{weight1}
\displaystyle w_1^i = \left\{\!\!\begin{array}{ll}
1, & \text{if } \frac{[|\boldsymbol\Psi(\hat{\mathbf x} - \mathbf x_0)|]_i}{1 + [|\boldsymbol\Psi(\hat{\mathbf x} - \mathbf x_0)|]_i} > \epsilon_1 \\
\frac1{1 + [|\boldsymbol\Psi\mathbf x_0|]_i}, & \mathrm{otherwise}
\end{array}
\right.
\end{equation}

\begin{equation} 
\label{weight2}
w_2^i = \frac{1}{1+ [|\hat{\bx} -\bx_0|]_i}
\end{equation}

where $\epsilon_1>0$ is some fixed threshold parameter.

In \eqref{optproblem}, term 1 enforces the sparsity constraint, while term 2 enforces the similarity of the $\hat{\bx}$ to the reference scan $\bx_0$.  Diagonal matrices $\bW_1$ and $\bW_2$ are iterative weighting terms that depend on the difference between the current reconstruction and the reference image.  Observe that if the difference between the $i$th component of $\hat{\bx}$ and $\bx_0$ is large, $w_1^i \approx 1$ and $w_2^i\approx 0$. In other words, when the difference between the reference image and the current scan is large, the optimization problem \eqref{optproblem} becomes conventional $\ell_1$ minimization with wavelet sparsity (L1-W):

\begin{equation}
\underset{\bx}{\text{min}} \, \normOne{\bPsi \bx}       
	\text { s.t. } \, \, \, \normTwo{ \bF_u \bx - \by } < \epsilon.
\label{L1-min}
\end{equation}

On the other hand, when $\hat{\bx}$ and $\bx_0$ are reasonably close, the $i$th entry of $\bW_1$ takes a large value to enforce sparsity if the $i$th entry of $\bPsi \bx_0$ is small, or relaxes the sparsity condition if the $i$th entry of $\bPsi \bx_0$ is large. $\bW_2$ behaves in a similar way. It enforces the similarity between $\hat{\bx}$ and $\bx_0$ for the regions that are highly similar while it relaxes the constraint for the regions that are not similar. $\lambda$ is a regularization parameter that controls the relative importance of term 1 and term 2.

In LACS-MRI, we select measurements using mixed PDFs of variable density and adaptive density functions.  In particular, we will utilize the following formal definitions of sampling densities:

\begin{equation}\label{fVD}
f_{VD}(k_y) = \frac{(1-\frac{2}{n}|k_Y|)^p}{\sum\limits_{k_y} (1-\frac{2}{n}|k_y|)^p}
\end{equation}

\begin{equation}\label{fND}
f_{ND}(k_y) = \frac{g_{ND}(k_y)}{\sum\limits_{k_y} g_{ND}(k_y)} ,\ \ \ \ \ \   g_{ND}(k_y) = \frac{\sum_{i \in k_y} [|\bF \hat{\bx} - \bF \bx|]_i}{[|\bF \hat{\bx}| + |\bF \bx|]_{i}}
\end{equation}

\begin{equation}\label{fR}
f_{R}(k_y) = \frac{g_{R}(k_y)}{\sum\limits_{k_y} g_{R}(k_y)} ,\ \ \ \ \ \   g_{R}(k_y) = \sum\limits_{i \in k_y} [|\bF \bx_0|]_i.
\end{equation}

We may then mix these PDFs as described in \eqref{mixedPDF}.  There, $\gamma$ is a fidelity term that controls the weight of variable and adaptive density sampling, where $\gamma = \frac{1}{N}\sum\limits_{i=1}^n w_{2}^i$.  This term increases when $\hat{\bx}$ and $\bx$ are similar, which leads to a heavier weighting of adaptive density sampling. Intuitively, it makes sense that we want to use the adaptive density sampling more than the variable density sampling when the reference image is close to the follow-up image since we expect most of the signal information to reside within reference image.

We will also consider other sampling strategies proposed in various works.  Krahmer et. al. suggest a nearly optimal Variable Density Sampling ($f_{VDS}$) function that is able to reconstruct images using various sparsification methods ~\cite{krahmer2014stable}. $f_{VDS}$ is a variable density sampling function that selects 2D $k$-space coordinate pairs ($k_1, k_2$) with probability:
\begin{equation}
\text{Prob} [(\omega^j_1, \omega^j_2) = (k_1, k_2)] = C' \, \text{min} \, \left ( C, \frac{1}{(k_1^2+k_2^2)^p}    \right ),\
\label{ward}
\end{equation} 
where $p$ is a user selected parameter (typically $0 < p < 2$), while $C$ and $C'$ denote appropriate constants.
Davenport et. al. suggest the optimal Constrained Adaptive Sensing ($f_A$) function to select measurements based on a signal estimation~\cite{davenport2015constrained}. $f_A$ is defined by taking a sequence of m rows from $\bF$, where $\bF$ is the discrete Fourier transform matrix (DFT) that transforms a vectorized image~\cite{davenport2015constrained}. Letting $\ba_i$ be the $i$th row of the discrete Fourier transform $\bF$, a sequence of $\ba_i$ can be selected by the following (where $\bA'$ denotes the submatrix with rows $\ba_i$):
\begin{equation}
\{\ba_i\}^m_{i=1} = \underset{\{\{\ba_i\}^m_{i=1}\}}{\arg\min} \mbox{     tr} (((\bA'\bPsi_\Lambda)*\bA'\bPsi_\Lambda)^{-1})
\end{equation}
However, this is intractable, so Davenport et. al. \cite{davenport2015constrained} consider the relaxation: 

\begin{equation}
\hat{\bS} = \underset{\mathrm{diagonal}\ \mathrm{matrices}\ \bS\succeq 0}{\arg\min} \mathrm{tr} (((\bF\bPsi_\Lambda)*\bS\bF\bPsi_\Lambda)^{-1})\hspace{0.50 cm} \mathrm{subject}\ \mathrm{to}\ \mathrm{tr}(\bS)\leq \varepsilon,
\end{equation}

where $\hat{\bS}$ is a diagonal matrix whose diagonal entries correspond to how much of the corresponding Fourier measurement should be utilized. Then, the probability of selecting the $i$th row of $\bF$ can be calculated by:  
\begin{equation}
\label{DMNW}
f_A(i) = \frac{\hat{S}_{ii}}{m}.
\end{equation}

The pseudocode for LACS-MRI as presented in ~\cite{weizman2014application} is given below.  We next turn to our numerical investigations of this method combined with various sampling techniques. 

\begin{algorithm}[H]
\caption{LACS-MRI}%\label{scale algorithm}
\begin{algorithmic}

\STATE \textbf{Input: }$\bx_0$ (the reference image), $\boldsymbol{N}$ (the number of iterations), $\eta$ (total percentage $k$-space to be sampled) \\%,  $\bF_u$, sub-sampled Fourier operator \\ 
\textbf{Output: } $\hat{\bx}$  \text{(estimated image)}

\STATE Let $\bW_1=I, \bW_2=0, \text{and } \by=0$.
Pick $k=\frac{\eta}{N}$ sampling locations randomly (using VDS or uniformly).
\\ \text{Sampling:} 
 \indent for i=1 to $\boldsymbol{N}$
 \STATE\hspace{\algorithmicindent} Sample $k$ locations, $\by_i=(\bF_u)_i\bX$  
 \STATE\hspace{\algorithmicindent}  Let $\by=\by+\by_i$
 \STATE \hspace{\algorithmicindent} Solve $\hat{\bx}= \underset{\bx}{\text{min}} \, \normOne{\bW_1 \bPsi \bx} + \lambda \normOne{ \bW_2 (\bx - \bx_0)} + \normTwo{ \bF_u \bx - \by } $
 \STATE \hspace{\algorithmicindent} Update $\bW_1$ and $\bW_2$ via \eqref{weight1} and \eqref{weight2}. 
 Use the sampling PDF \eqref{mixedPDF} to select $k$ new sampling locations. 

\end{algorithmic}
\end{algorithm}

\section{Experimental Results}

\subsection{Reconstruction performance with various PDFs}
We compare the image recovery of $\ell_1$ minimization with wavelet sparsity (L1-W) and Adaptive Weighted Temporal CS-MRI (LACS-MRI) algorithms through experiments. L1-W is a conventional compressed sensing problem given in \eqref{L1-min}.  
We solve this problem using the Projection Over Convex Sets (POCS) method, which was shown to be faster and to have a better recovery in preliminary experiments \cite{lustig2010spirit}.
Weizman et.al. have tested the LACS-MRI algorithm with Variable Density Sampling for LACS-MRI ($f_{VD}$), Reference LACS-MRI sampling ($f_{ND}$), and Adaptive Density Sampling for Reference ($f_{R}$) ~\cite{weizman2014application}. However, these sampling functions have not been proven to be optimal. These algorithms can be improved by utilizing other variable density or adaptive density sampling functions. 

We consider twenty two combinations of experiments in Table \ref{CaseReference} and measure their mean Reconstruction Signal to Noise Ratios (RSNR) over 50 iterations to decide the best scheme.
The mixture of variable density sampling and adaptive density sampling is defined by \eqref{mixedPDF} with a fidelity term $\gamma$ to select measurements in $k$-space in the LACS-MRI method.
The idea of creating a mixed PDF can be applied to classical L1-minimization as well. Thus, we experimented with both LACS-MRI and classical L1-minimization (L1-W) along with two variable density sampling PDFs, $f_{VD}$ and $f_{VDS}$, and three different adaptive density sampling PDFs, $f_{R}$, $f_A$, and $f_{ND}$ to create mixed PDFs as in \eqref{mixedPDF}. These PDFs are described in \eqref{fVD}, \eqref{ward}, \eqref{fR}, \eqref{DMNW}, and \eqref{fND}, respectively.
For some of the tests, we purely used adaptive density sampling or variable density sampling. For each unique PDF defined, we experimented with a 32 x 32 phantom and a 512 x 512 brain image according to Table \ref{CaseReference}. For each true image $\bx_\star$ and recovered estimation $\hat{\bx}$, we calculated the Reconstruction SNR defined by RSNR = $20\log(\normTwo{\bx_\star} / \normTwo{\bx_\star - \hat{\bx}})$, over 50 trials and calculated the mean over all 50 trials (MSNR). We also varied the compression level $\eta = \frac{m}{n}$, where $m$ is the number of measurements, and $n$ is the dimension of the image. We tested each of the cases, varying $\eta$ from 0.03-0.21 in increments of 0.03 to observe the dynamics of the MSNR.

\begin{table}[ht]
\begin{center}
\caption{Case Reference Table}
\label{CaseReference}
\begin{tabular}{llll}
\multicolumn{4}{c}{} \\
\cline{1-4}
\textit{Cases}    & \textit{Variable Density Sampling} & \textit{Adaptive Density Sampling} & \textit{ Algorithm} \\
\hline
case 1  & $f_{VD}$ & $f_{R}$ & LACS-MRI   \\
case 2  & $f_{VDS}$   & $f_{R}$ & LACS-MRI   \\
case 3  & $f_{VD}$   & - & LACS-MRI   \\ 
case 4  & $f_{VDS}$ & - & LACS-MRI   \\
case 5 & -    & $f_{R}$ & LACS-MRI   \\
case 6 & -    & $f_A$ & LACS-MRI   \\
case 7 &$f_{VD}$    & $f_A$ & LACS-MRI   \\
case 8 & $f_{VDS}$   & $f_A$ & LACS-MRI   \\
case 9  & $f_{VD}$    & $f_{R}$ & L1-W   \\
case 10  & $f_{VDS}$    & $f_{R}$ & L1-W   \\
case 11  & $f_{VD}$   & - & L1-W   \\ 
case 12  & $f_{VDS}$   & - & L1-W   \\
case 13 & -    & $f_{R}$ & L1-W   \\
case 14 & -    & $f_A$ & L1-W   \\
case 15 & $f_{VD}$    & $f_A$ & L1-W   \\
case 16 & $f_{VDS}$    & $f_A$ & L1-W   \\
case 17 & $f_{VD}$    & $f_{ND}$& LACS-MRI   \\
case 18 & $f_{VDS}$    & $f_{ND}$ & LACS-MRI   \\
case 19 & -    & $f_{ND}$ & LACS-MRI   \\
case 20 & $f_{VD}$    & $f_{ND}$ & L1-W   \\
case 21 & $f_{VDS}$    & $f_{ND}$ & L1-W   \\
case 22 & -    & $f_{ND}$ & L1-W   \\
\end{tabular}
\end{center}
\end{table}
%%%%%%  Phantom figures %%%%%%%%%%%%%
\begin{figure}[ht]
\begin{subfigure}{0.24\textwidth}
\centering
\includegraphics[scale=0.35]{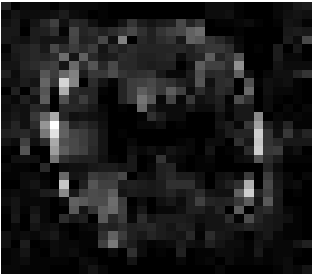}
% in order to find the original '.fig' file, go to finalfigs/BriFigs
\caption{$\eta$ = 0.06}
%\label{PLmin6}
\end{subfigure} 
\begin{subfigure}{0.24\textwidth}
\centering
\includegraphics[scale=0.35]{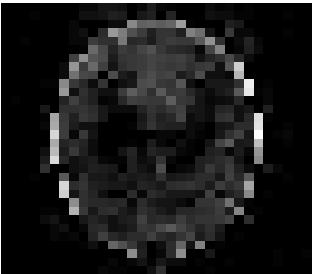}
% in order to find the original '.fig' file, go to finalfigs/BriFigs
\caption{$\eta$ = 0.12}
%\label{PLmin12}
\end{subfigure}
\begin{subfigure}{0.24\textwidth}
\centering
\includegraphics[scale=0.35]{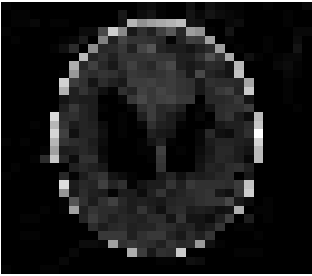}
% in order to find the original '.fig' file, go to finalfigs/BriFigs
\caption{$\eta$ = 0.18}
%\label{PLmin18}
\end{subfigure}
\begin{subfigure}{0.24\textwidth}
\centering
\includegraphics[scale=0.35]{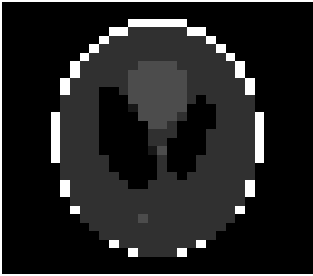}
% in order to find the original '.fig' file, go to finalfigs/BriFigs
\caption{Ground Truth/Follow-up}
%\label{Pground}
\end{subfigure}
\begin{subfigure}{0.24\textwidth}
\centering
\includegraphics[scale=0.35]{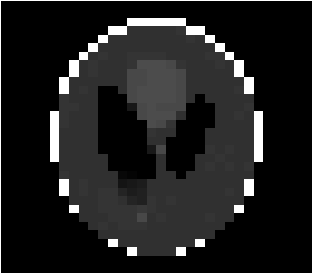}
% in order to find the original '.fig' file, go to finalfigs/BriFigs
\caption{$\eta$ = 0.06}
%\label{PLACS6}
\end{subfigure} 
\begin{subfigure}{0.24\textwidth}
\centering
\includegraphics[scale=0.35]{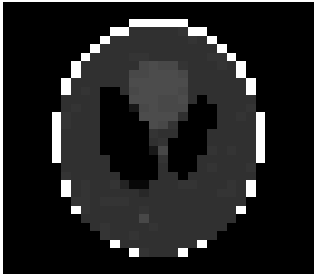}
% in order to find the original '.fig' file, go to finalfigs/BriFigs
\caption{$\eta$ = 0.12}
%\label{PLACS12}
\end{subfigure}
\begin{subfigure}{0.24\textwidth}
\centering
\includegraphics[scale=0.35]{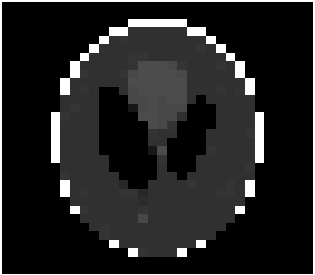}
% in order to find the original '.fig' file, go to finalfigs/BriFigs
\caption{$\eta$ = 0.18}
%\label{PLACS18}
\end{subfigure}
\begin{subfigure}{0.24\textwidth}
\centering
\includegraphics[scale=0.35]{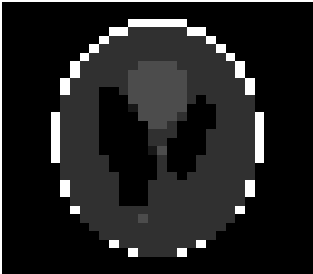}
% in order to find the original '.fig' file, go to finalfigs/BriFigs
\caption{Reference}
\label{Preference}
\end{subfigure}
%\label{phantom}
\caption{Image reconstruction from case 13 (a, b, and c) and case 1 (e, f, and g) of a phantom image. Case 13 is the best performing algorithm that uses L1-W and case 1 is the best performing algorithm that uses LACS-MRI}\label{phantomfigs}
\end{figure}
%%%%%% end of Phantom figures %%%%%%%%%%%%%
\FloatBarrier

\begin{figure}[ht]
\begin{subfigure}{0.24\textwidth}
\centering
\includegraphics[scale=0.35]{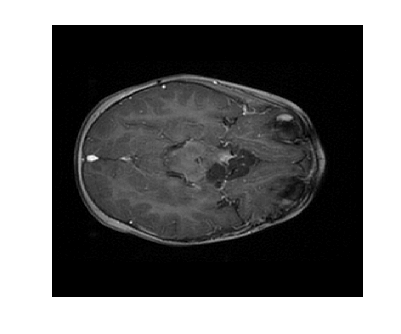}
% in order to find the original '.fig' file, go to finalfigs/BriFigs
\caption{$\eta$ = 0.06}
%\label{PLmin6}
\end{subfigure} 
\begin{subfigure}{0.24\textwidth}
\centering
\includegraphics[scale=0.35]{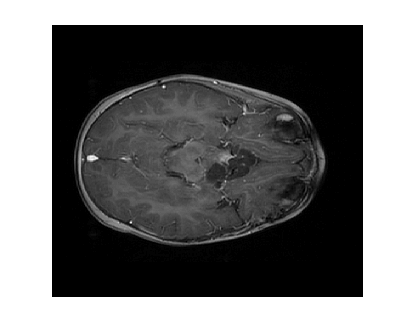}
% in order to find the original '.fig' file, go to finalfigs/BriFigs
\caption{$\eta$ = 0.12}
%\label{PLmin12}
\end{subfigure}
\begin{subfigure}{0.24\textwidth}
\centering
\includegraphics[scale=0.35]{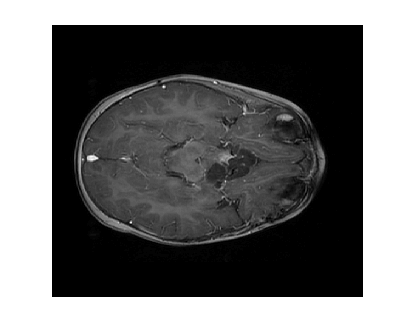}
% in order to find the original '.fig' file, go to finalfigs/BriFigs
\caption{$\eta$ = 0.18}
%\label{PLmin18}
\end{subfigure}
\begin{subfigure}{0.24\textwidth}
\centering
\includegraphics[scale=0.35]{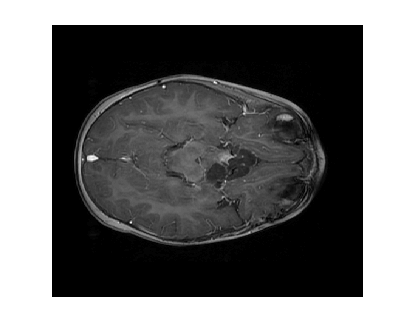}
% in order to find the original '.fig' file, go to finalfigs/BriFigs
\caption{Ground Truth/Follow-up}
%\label{Pground}
\end{subfigure}
\begin{subfigure}{0.24\textwidth}
\centering
\includegraphics[scale=0.35]{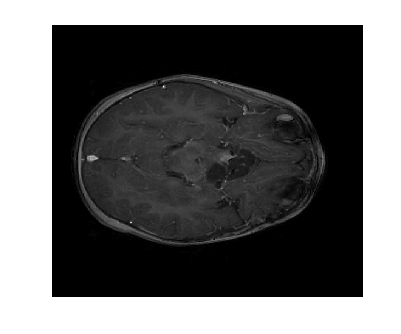}
% in order to find the original '.fig' file, go to finalfigs/BriFigs
\caption{$\eta$ = 0.06}
%\label{PLACS6}
\end{subfigure} 
\begin{subfigure}{0.24\textwidth}
\centering
\includegraphics[scale=0.35]{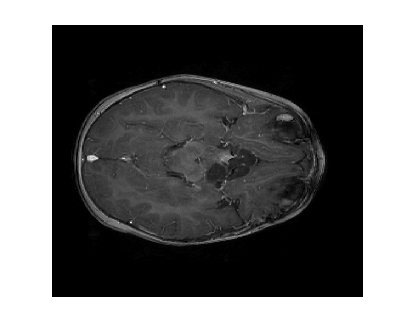}
% in order to find the original '.fig' file, go to finalfigs/BriFigs
\caption{$\eta$ = 0.12}
%\label{PLACS12}
\end{subfigure}
\begin{subfigure}{0.24\textwidth}
\centering
\includegraphics[scale=0.35]{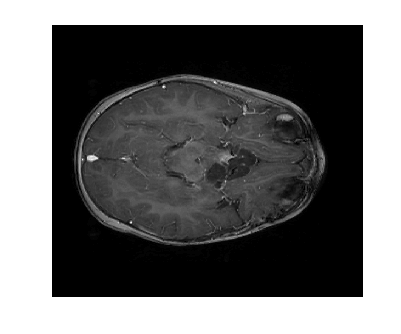}
% in order to find the original '.fig' file, go to finalfigs/BriFigs
\caption{$\eta$ = 0.18}
%\label{PLACS18}
\end{subfigure}
\begin{subfigure}{0.24\textwidth}
\centering
\includegraphics[scale=0.35]{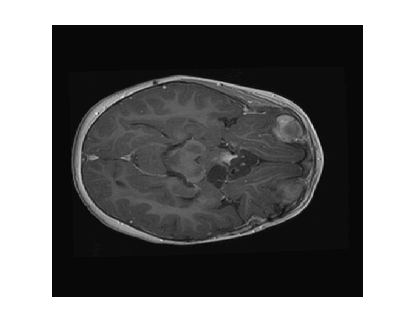}
% in order to find the original '.fig' file, go to finalfigs/BriFigs
\caption{Reference}
\label{Breference}
\end{subfigure}
%\label{phantom}
\caption{Image reconstruction from Case 20 (a, b, and c) and Case 3 (e, f, and g) of a brain image. Case 20 is the best performing algorithm that uses L1-W and case 3 is the best performing algorithm that uses LACS-MRI}\label{brainfigs}
\end{figure}

\begin{figure}[ht]
\begin{subfigure}{0.49\textwidth}
\centering
\includegraphics[scale=0.45]{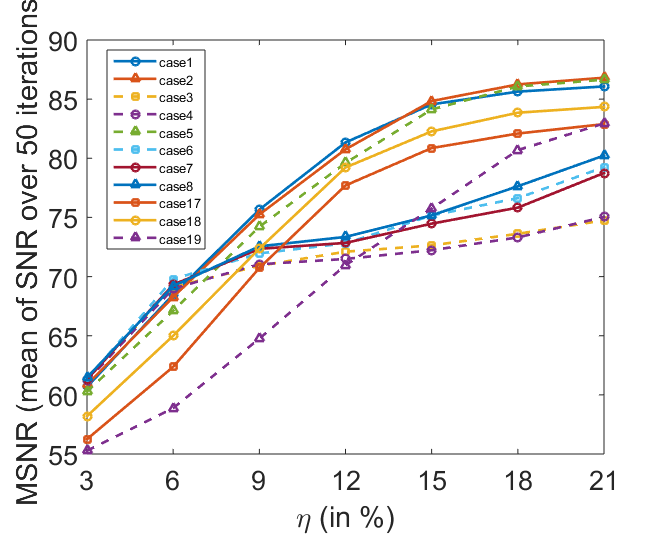}
% in order to find the original '.fig' file, go to finalfigs/BriFigs
\caption{Phantom image with high MSNR cases}
\label{phantomHigh}
\end{subfigure} 
\begin{subfigure}{0.49\textwidth}
\centering
\includegraphics[scale=0.45]{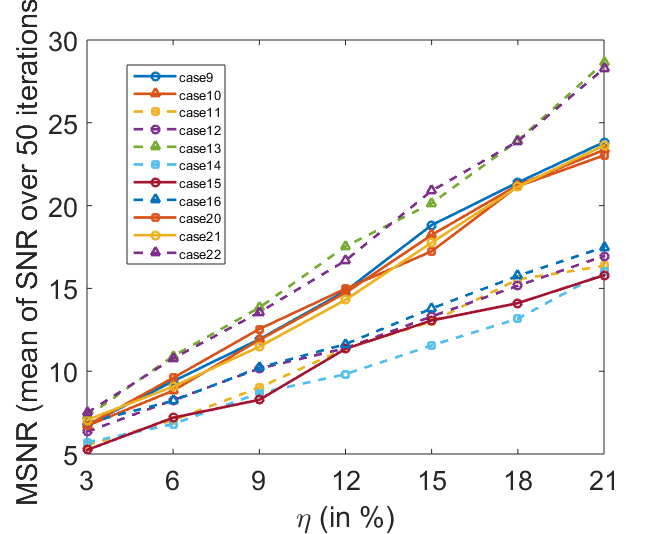}
% in order to find the original '.fig' file, go to finalfigs/BriFigs
\caption{Phantom image with low MSNR cases}
\label{phantomLow}
\end{subfigure}
\begin{subfigure}{0.49\textwidth}
\centering
\includegraphics[scale=0.45]{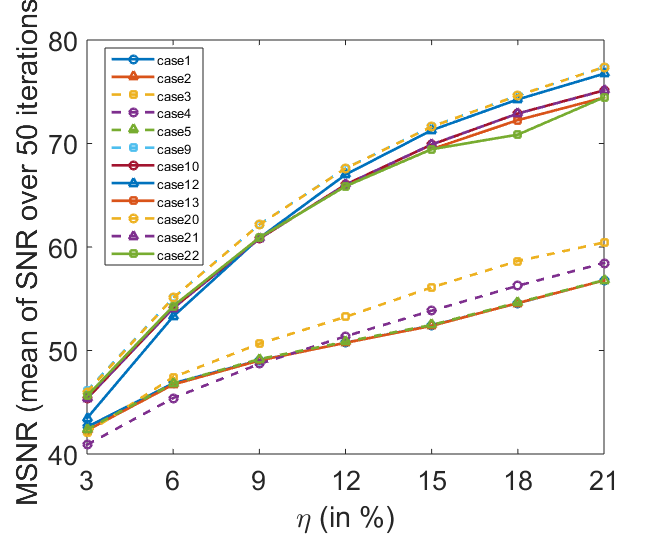}
% in order to find the original '.fig' file, go to finalfigs/BriFigs
\caption{Brain image with high MSNR cases}
\label{brainHigh}
\end{subfigure} 
\begin{subfigure}{0.49\textwidth}
\centering
\includegraphics[scale=0.45]{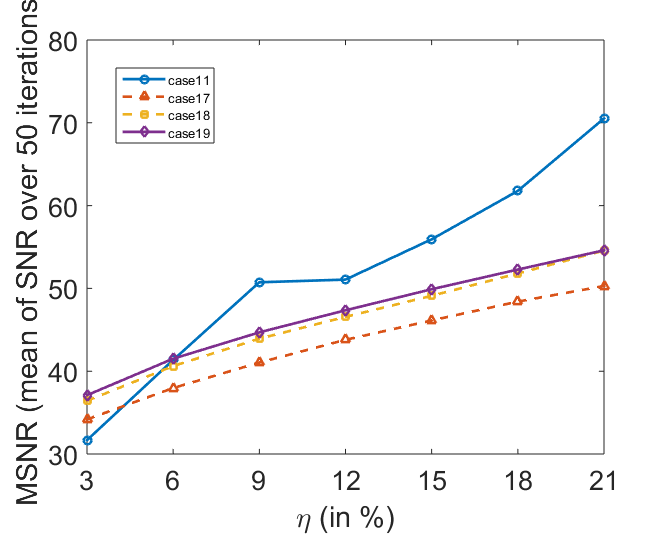}
% in order to find the original '.fig' file, go to finalfigs/BriFigs
\caption{Brain image with low MSNR cases}
\label{BrainLow}
\end{subfigure}
\caption{Summary of Phantom and Brain image experiments}
\label{SimSummary}
\end{figure}

Figure \ref{SimSummary} displays the reconstruction MSNR for the various test cases as cataloged in Table \ref{CaseReference}.  Figures \ref{phantomfigs} and \ref{brainfigs} display some examples of reconstructed images for cases of interest.

The image recovery results for the phantom image experiment were quite extreme, as shown in Figure \ref{SimSummary}. We split all the cases into two groups: a group with high MSNR and a group with low MSNR. While grouping these cases we noticed that all of the cases that used LACS-MRI were classified as high MSNR, outperforming the cases that used L1-W.  
This is expected since L1-W is unable to utilize a reference image.  Case 1 and Case 2 had the best recovery of the phantom image out of all methods. Case 5, which also uses the same adaptive sampling density as cases 1 and  2, also shows accurate recovery. We notice that for cases with $f_R$ (Case 1, Case 2, and Case 5) and $f_{ND}$ (Case 17, Case 18, and Case 19) as their adaptive density sampling methods, the RSNR is high when using the LACS-MRI algorithm. $f_{VD}$ combined with $f_{R}$ performs better than $f_{VDS}$ combined with $f_R$. Accordingly, we see that the adaptive density function plays a key role in the LACS-MRI algorithm. 

\indent Note that when we use only variable density sampling, we get the worst performance. Again, this is not surprising since this means we do not benefit from the use of a reference image. Recovery can be improved by using $f_A$ but it  does not compare favorably with $f_R$ or $f_{ND}$. On the other hand,  L1-W recovered the image poorly. Surprisingly, purely adaptive cases with $f_{R}$ and $f_{ND}$ (Case 13 and Case 14) performed the best among all  combinations. In L1-W, we initialize $f_{ND}$ to be the same as $f_{R}$ which is why both show almost the exact same performance. 

%%%%%%  Brain figures %%%%%%%%%%%%%
%%%%%% end of Phantom figures %%%%%%%%%%%%%
\FloatBarrier
The phantom image that we used in the previous experiment is flat and has less texture while the brain images has a lot of wrinkles and complex shapes. While the phantom image is small, the brain image is too large to calculate $f_{ND}$ efficiently. Thus, we conducted an experiment in the same setting as the phantom image excluding cases using $f_A$: cases 6 through 8 and 14 through 16.
In this experiment, contrary to results for the phantom image experiment, L1-W performs better than LACS-MRI. For example, Case 3 performed the best among cases using LACS-MRI but it performs worse than 6 cases that use L1-W. Interestingly, one of the cases that shows an outstanding image recovery purely uses $f_{VDS}$ with L1-W. However, purely using $f_{VD}$ with L1-W shows a dramatically worse performance.  
In conclusion, we are unsure why each combination of mixed PDFs and reconstruction algorithm perform differently based on the images processed, but this differing behavior is crucial to highlight for practical applications.  We conjecture that the texture of the image is one factor that decides the performance of each algorithm. In the future, we may explore this conjecture and analyze why this happens.
%%%%%%% summary plot%%%%%%%%%%%%%%%%%%%%%
% This summary plot can be generated by 'b' in Final Report/BriCode/plot_simulation. Please run the script if you want to generate this figures. 
%%%%%%% end of summary plot%%%%%%%%%%%%%%%%%%%%%
\FloatBarrier

\subsection{Optimal parameters for sampling probability density functions}

We next present simulations in which we tested different values of $p$ and $C$ in Krahmer and Ward's~\cite{krahmer2014stable} variable density sampling function, which we will again refer to here as $f_{VDS}$. Recall $f_{VDS}$ is defined via 

\begin{align*}
	\text{Prob} [(\omega^j_1, \omega^j_2) = (k_1, k_2)] = C' \, \text{min} \, \left ( C, \frac{1}{(k_1^2+k_2^2)^p}    \right ),
\end{align*}
where $\omega^j_i$ are frequencies, $k_i$ are $k$-space coordinates, and $C'$, $C$, and $p$ are constants. 
In our first experiment, we measured RSNR for $.02 \le p \le 1.5$ at increments of .02, and $C = 1, \, .1, \, .01$, and .001, in an attempt to find optimal values of $C$; all experiments were performed on the brain image as above. We used the LACS-MRI algorithm to reconstruct the image and, in keeping with Krahmer and Ward's methods, set the compression level $\eta = .06$. %\ref{Low_SNR_Cases}

\begin{figure}[ht]
\centering
 \includegraphics[scale=.45]{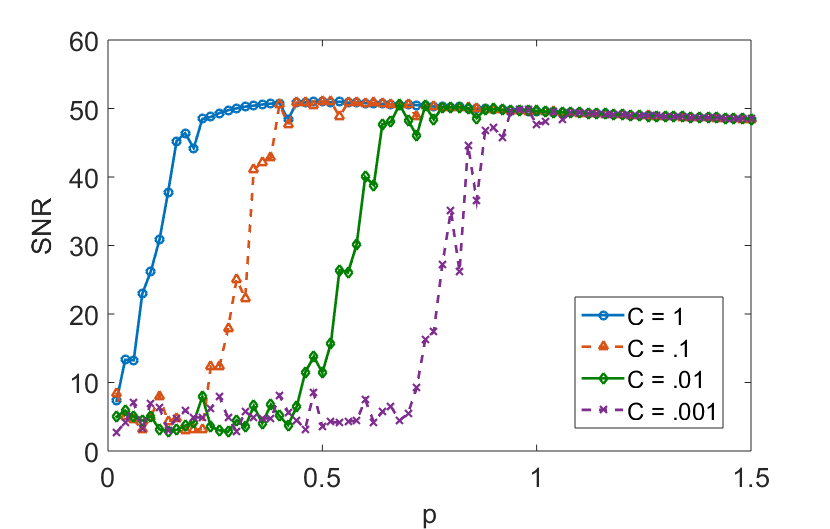}
\caption{Reconstruction (by LACS-MRI) RSNR using $f_{VDS}$ with different values of $C$ and $p$.} \label{figCs}
%\label{Low_SNR_Cases}
\end{figure}
\FloatBarrier

The results are shown in Figure \ref{figCs}.  We found that the larger $C$ is, the greater the RSNR of the reconstruction is. However, as $p$ becomes sufficiently large, there is virtually no difference in performance among the difference values of $C$. This is entirely consistent with how $f_{VDS}$ is defined. An intuitive way to visualize $f_{VDS}$ is to imagine the three-dimensional graph of $\frac{1}{(k_1^2+k_2^2)^p}$ where $k_1$ and $k_2$ are independent and $p$ is fixed. The effect of selecting the smaller value of $C$ and $\frac{1}{(k_1^2+k_2^2)^p}$ is to place an upper bound on $\frac{1}{(k_1^2+k_2^2)^p}$, which would otherwise blow up near the origin. $f_{VDS}$ is precisely the two-dimensional intersection of $\frac{1}{(k_1^2+k_2^2)^p}$ and the plane $z = C$. In our earlier discussion about variable density sampling, we argue that sampling primarily near the origin in $k$-space yields a superior reconstruction. Conversely, a probability density function that samples uniformly at random everywhere will perform poorly. When $p$ is small, $\frac{1}{(k_1^2+k_2^2)^p}$ is not very steep, even as $(k_1, k_2) \rightarrow 0$. Graphically, $f_{VDS}$ has a large diameter and therefore will produce a lower RSNR image reconstruction. Furthermore, since $\frac{1}{(k_1^2+k_2^2)^p}$ is not steep, the difference in the probability density function for each value of $C$ is drastic, which explains why higher values of $C$ perform much better than lower values of $C$ for low $p$. However, when $p$ is large, $\frac{1}{(k_1^2+k_2^2)^p}$ becomes very steep near the origin, which results in both a small probability density function and a similar performance for each $C$. 

Next, we tested the effects on RSNR by varying $p$ while using mixed PDFs with $f_{VDS}$.  We used different adaptive sampling functions and reconstruction algorithms.  Specifically, we tested the LACS-MRI and $\ell_{1}$-minimization with wavelet sparsity (L1-W) reconstruction algorithms with the adaptive-weighted reference based MRI ($f_R$), $f_{VDS}$ only, $f_A$, and $f_{VDS}$ and $f_{ND}$ adaptive sampling functions. As discussed previously, LACS-MRI delivers an image reconstruction that is significantly superior to L1-W. We separated these eight cases into two graphs based on whether they were reconstructed using LACS-MRI or $L_1$-W, and thus whether they yielded high or low RSNR. 

\begin{figure}[!ht]
\begin{subfigure}{0.49\textwidth}
\centering
\includegraphics[scale=0.35]{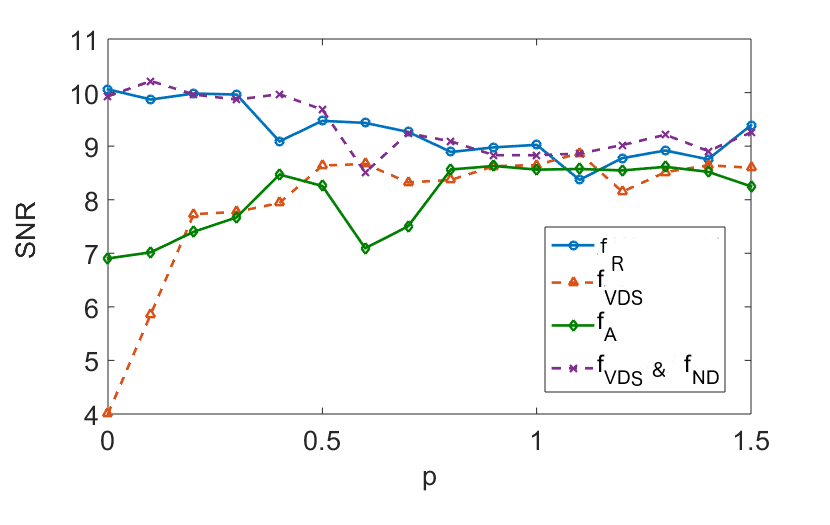}
%\caption{Reconstruction SNR as a function of $p$ using L1-W.}
%\label{Low_SNR_Cases}
\end{subfigure} 
\begin{subfigure}{0.49\textwidth}
\centering
\includegraphics[scale=0.35]{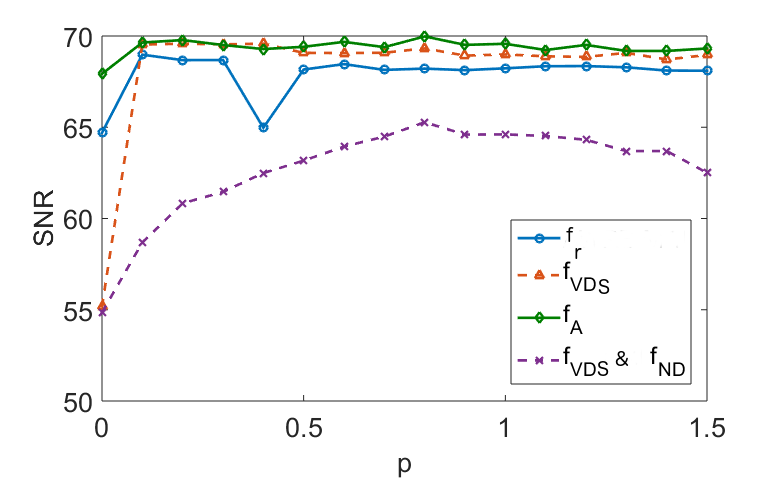}
%\caption{Reconstruction SNR as a function of $p$ using LACS-MRI.}
%\label{High_SNR_Cases}
\end{subfigure}
%\label{p vs SNR graphs}
\caption{Left: Reconstruction SNR as a function of $p$ using L1-W. Right: Reconstruction SNR as a function of $p$ using LACS-MRI.}\label{mixedp}

\end{figure}
\begin{figure}[ht]
\includegraphics[scale=0.33]{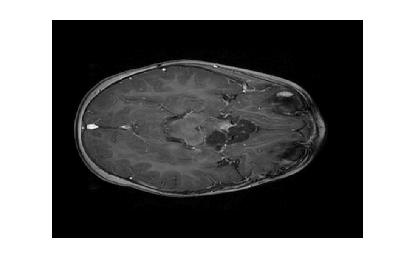} \includegraphics[scale=0.33]{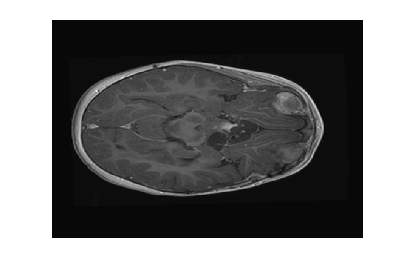}
\includegraphics[scale=0.33]{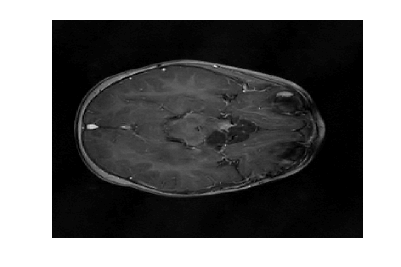} \includegraphics[scale=0.33]{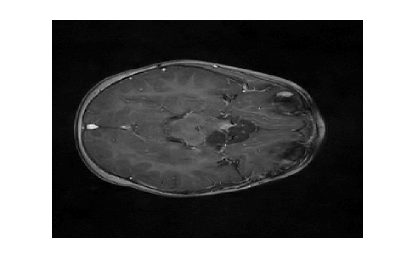} \includegraphics[scale=0.33]{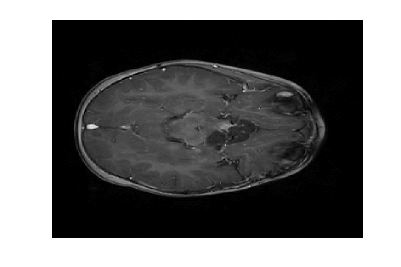} 
\caption{From left to right: the ground truth, reference image, and reconstructions  for optimal $p$, $p=1$ and $p=1.5$. }\label{pbrain}
\end{figure}

The results of this experiment are shown in Figure \ref{mixedp}.
In both the high RSNR and low RSNR cases, all combinations followed the same pattern as $p$ varied, suggesting that the value of $p$ is a significant factor in all reconstructions that utilize $f_{VDS}$. None of the low RSNR cases showed any discernible patterns as $p$ varied. We conjecture that this is because the RSNR is simply too low to draw any meaningful results. On the other hand, reconstructions of the high RSNR cases do give meaningful results. First, although theoretical results of $f_{VDS}$ suggest that the optimal value of $p$ should be close to 1.5, we see here that instead the optimal value of $p$ is around .7.  However, also note that there is very little change in RSNR from the optimal $p$ value and all other $p$ values in a relatively large range. We conjecture that if $p$ is within a sufficiently large range such as $.5 \le p \le 1.5$, the RSNR of the reconstruction will be bounded below, guaranteeing an accurate reconstruction.  Examples of resconstructed images are shown in Figure \ref{pbrain}, where there are only very minor discernable differences in reconstruction.

\subsection{Total Variation and gradient sparsification}

Total variation (TV) minimization is a common process in image processing used for denoising and other applications. The \textit{total variation} of an image $\bX$ refers to the $\ell_{1}$ norm of the discrete gradient, $\| \boldsymbol{X} \|_{TV} = \| \nabla \boldsymbol{X} \|_{1}$, where $\nabla$ denotes the discrete gradient operator that concatenates the horizontal pixel-wise differences and the vertical differences.  
Suppose we have an $N \mathrm{x} N$ image, $\boldsymbol{X} \in \mathbb{C}^{NxN}$. Let $\boldsymbol{X}_{j,k}$ be the pixel in the $j^{th}$ row and $k^{th}$ column: 
\begin{equation}
\boldsymbol{X} = \begin{bmatrix}
    x_{11} & x_{12} & \dots & x_{1N} \\
    x_{21} & x_{22} & \dots & x_{2N} \\
    \vdots & \vdots & \ddots & \vdots \\
    x_{N1} & x_{N2} & \dots  & x_{NN}
\end{bmatrix}.
\end{equation}

\hspace{0.50cm} We can define both the vertical and horizontal pixel-wise change by taking the respective directional derivatives:
\begin{equation}
(\boldsymbol{X}_{x})_{j,k} = \left \{ \begin{tabular}{cc} 
$\boldsymbol{X}_{j,k} - \boldsymbol{X}_{j+1,k}$ & \hspace{0.50cm} $ 1 \leq j \leq N \ , \ 1 \leq k < N-1 $ \\ 
$\boldsymbol{X}_{j,k}$ & \hspace{0.50cm} $1 \leq j \leq N \ , \ k = N $
\end{tabular} \right. 
\label{X_Grad}
\end{equation}

\begin{equation}
(\boldsymbol{X}_{y})_{j,k} = \left \{ \begin{tabular}{cc}
$\boldsymbol{X}_{j,k} - \boldsymbol{X}_{j,k+1}$ & \hspace{0.50cm} $ 1 \leq j < N-1 \ , \ 1 \leq k \leq N $ \\
$\boldsymbol{X}_{j,k}$ & \hspace{0.50cm} $ j = N \ , \ 1 \leq k \leq N $
\end{tabular} \right. 
\label{Y_Grad}
\end{equation}

\hspace{0.50cm} For example, we can write the vertical discrete gradient transform as a matrix in the form:
\begin{equation}
\boldsymbol{G} = \begin{bmatrix}
    1 & -1 & 0 & \dots & \dots & 0 \\
    0 & 1 & -1 & 0 & \dots & 0 \\
    \vdots & \ddots & \ddots & \ddots & \ddots & \vdots \\
    \vdots & & \ddots & \ddots & \ddots & 0 \\
    \vdots & & & \ddots & \ddots & -1 \\
    0 & \dots & \dots & \dots & 0 & 1 
\end{bmatrix},
\label{Grad_Transform}
\end{equation}
such that the vertical and horizontal gradients of $\boldsymbol{X}$ are \(\boldsymbol{\nabla}_{y} = \boldsymbol{G}\boldsymbol{X} \ \mathrm{and} \ \boldsymbol{\nabla}_{x} = \boldsymbol{G}\boldsymbol{X}^{T}\).  We then write $\nabla\bX = (\boldsymbol{\nabla}_{x}, \boldsymbol{\nabla}_{y})$.

\begin{figure}[ht]

        \includegraphics[scale=0.28]{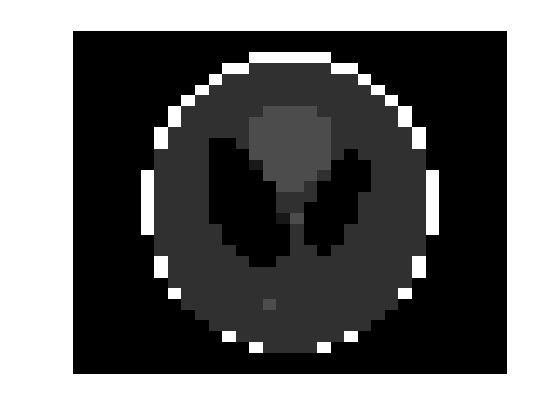} 
        \includegraphics[scale=0.28]{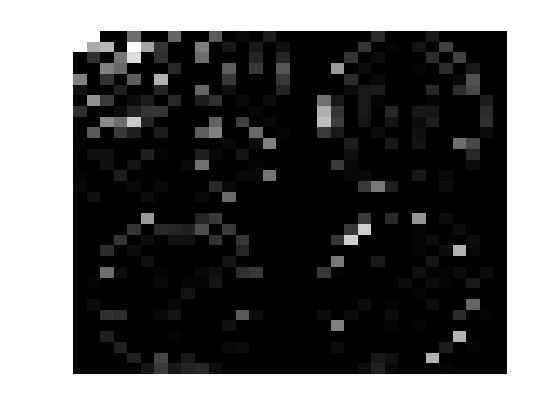}  
        \includegraphics[scale=0.28]{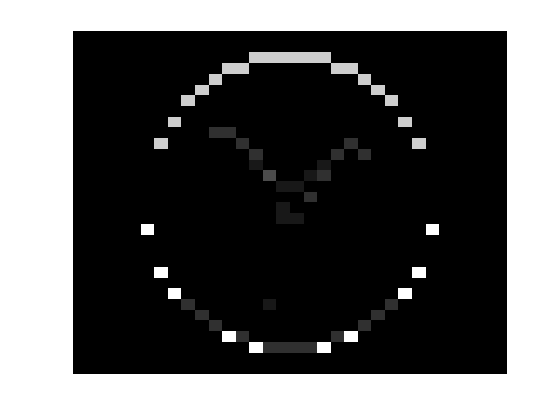}   
        \includegraphics[scale=0.3]{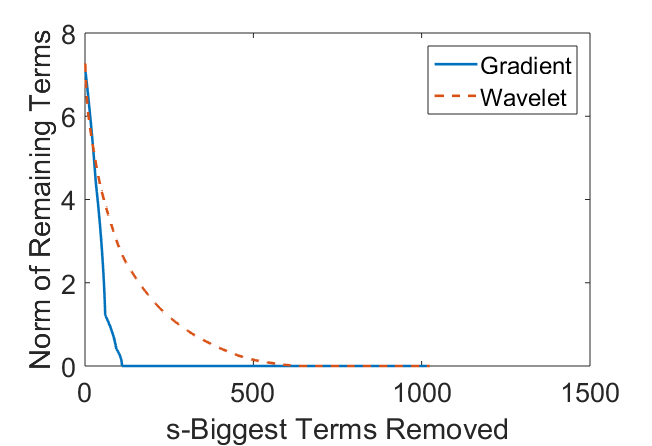}

    \caption{From left to right: original 32 x 32 Shepp-Logan Phantom image, wavelet representation, discrete gradient, and plot showing the decay of the Frobenius norm of the gradient/wavelet representation after removing large terms.}
    \label{P_Sparsity}
\end{figure}

When the total variation of an image is minimized, noise is removed and texture is smoothed out while the edges and important details are maintained. 
In the context of compressed sensing, the discrete gradient can be used as a sparsifying basis.  It was proven by Needell and Ward ~\cite{RefWorks:7} that even with undersampling, total variation minimization can recover an image robustly.  Indeed, many natural images do not vary much in intensity from pixel to pixel which makes them compressible with respect to their discrete gradients. 
	For example, the Shepp-Logan Phantom image is sparser with respect to its discrete gradient than its wavelet representation, see Figure \ref{P_Sparsity}. 

%%% Figure: Phantom With Different Sparsifying Transforms %%%
\FloatBarrier

\begin{figure}[ht]
    \centering
    \begin{subfigure}[h]{0.3\textwidth}
        \centering
        \includegraphics[scale=0.35]{Phantom.png}
        % Final Report >> finalfigs >> StephanieFigures >> Phantom.png/fig/mat
        \caption{}
    \end{subfigure}
    \begin{subfigure}[h]{0.3\textwidth}
        \centering
        \includegraphics[scale=0.35]{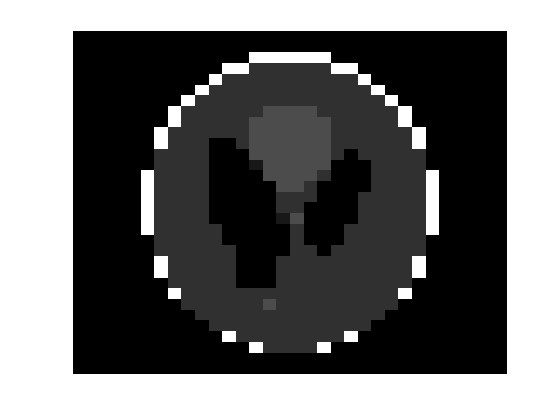}
        % Final Report >> finalfigs >> StephanieFigures >> Phantom_follow_up.png/fig/mat
        \caption{}
    \end{subfigure}
    \caption{(a) Reference phantom image and (b) Follow-up phantom image. Note the difference in the bottom of the left black spot; a "tumor" added  by changing a few pixel intensities.}
    \label{P_Ref_Follow}
\end{figure}

 Due to its wide use in image processing, we consider the use of gradient sparsification with LACS-MRI.  We show below that it is possible to produce image reconstructions with significantly higher Reconstruction SNR using gradient sparsification. We have repeated some of the simulations from above for different compression levels, $\eta$, using the gradient transform to sparsify instead of the wavelet transform.  We will refer to this as "LACS-MRI with gradient sparsity." We compared the performance of each sparsification method at each compression level $\eta$ using LACS-MRI and then compared this to the L1-W method. Again, we use a reference image and a follow-up image that we hope to recover as in Figure \ref{P_Ref_Follow}.  We consider here only the phantom image, as it serves as a prime candidate for gradient sparsification, and is small enough that evaluation of all PDFs is computationally feasible.

%%% Figure: Reference and Follow-Up Phantom %%%
\FloatBarrier

\begin{figure}[ht]
    \centering
    \begin{subfigure}[h]{0.3\textwidth}
        \centering
        \includegraphics[scale=0.3]{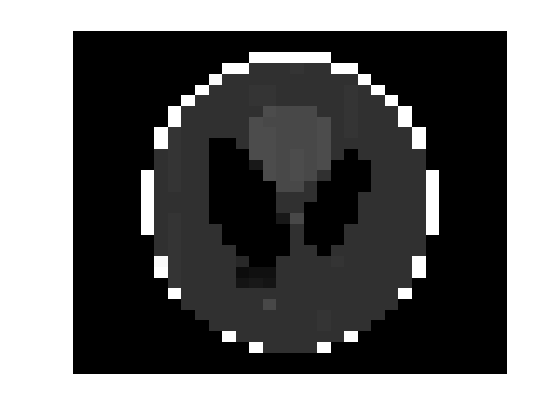}
        % Final Report >> finalfigs >> StephanieFigures >> Eldar_Eldar_Grad_6.png/fig/mat
        \caption{}
    \end{subfigure}
    \begin{subfigure}[h]{0.3\textwidth}
        \centering
        \includegraphics[scale=0.3]{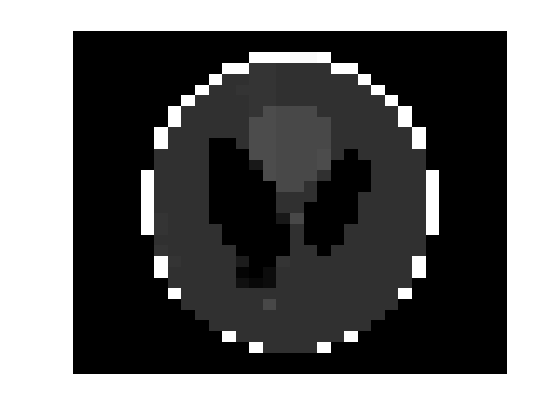}
        % Final Report >> finalfigs >> StephanieFigures >> Eldar_Eldar_Grad_12.png/fig/mat
        \caption{}
    \end{subfigure}
    \begin{subfigure}[h]{0.3\textwidth}
        \centering
        \includegraphics[scale=0.3]{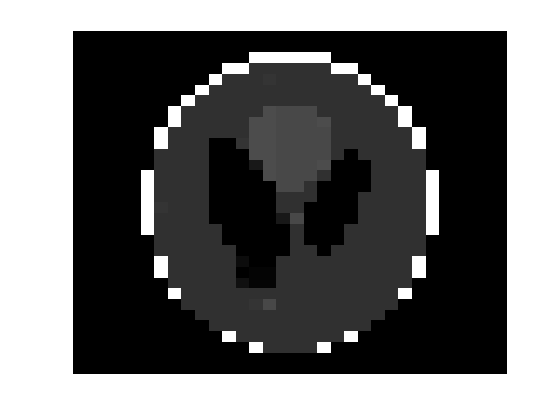}
        % Final Report >> finalfigs >> StephanieFigures >> Eldar_Eldar_Grad_18.png/fig/mat
        \caption{}
    \end{subfigure}
    \begin{subfigure}[h]{0.3\textwidth}
        \centering
        \includegraphics[scale=0.3]{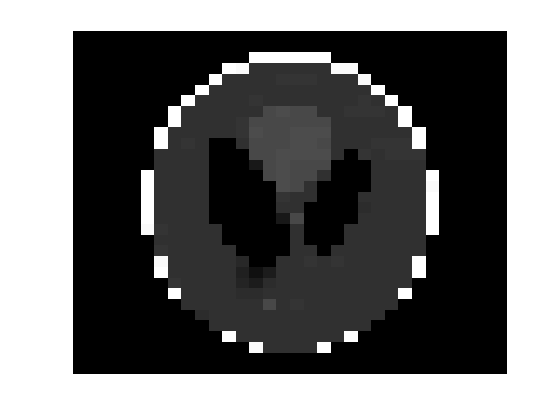}
        % Final Report >> finalfigs >> StephanieFigures >> Eldar_Eldar_Wave_6.png/fig/mat
        \caption{}
    \end{subfigure}
    \begin{subfigure}[h]{0.3\textwidth}
        \centering
        \includegraphics[scale=0.3]{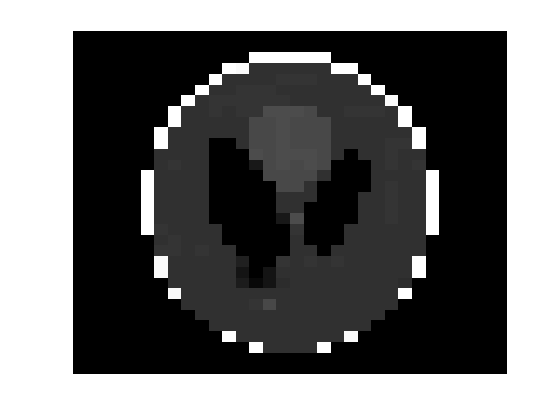}
        % Final Report >> finalfigs >> StephanieFigures >> Eldar_Eldar_Wave_12.png/fig/mat
        \caption{}
    \end{subfigure}
    \begin{subfigure}[h]{0.3\textwidth}
        \centering
        \includegraphics[scale=0.3]{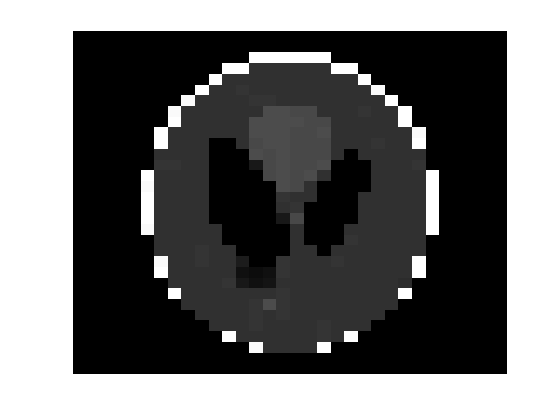}
         % Final Report >> finalfigs >> StephanieFigures >> Eldar_Eldar_Wave_18.png/fig/mat
        \caption{}
    \end{subfigure}
    \caption{(Top) Case 17 using LACS-MRI with gradient sparsification. Reconstruction with compression levels (a) 6 (b) 12 and (c) 18. (Bottom) Case 17 using LACS-MRI with wavelet sparsification. Reconstruction with compression levels (d) 6 (b) 12 and (c) 18. The sampling methods used together were $f_{VD}$ and $f_{ND}$.}
    \label{Phantom_Recovery}
\end{figure}

\begin{figure}[ht]
    \centering
    \begin{subfigure}[h]{0.4\textwidth}
        \centering
        \includegraphics[scale=0.3]{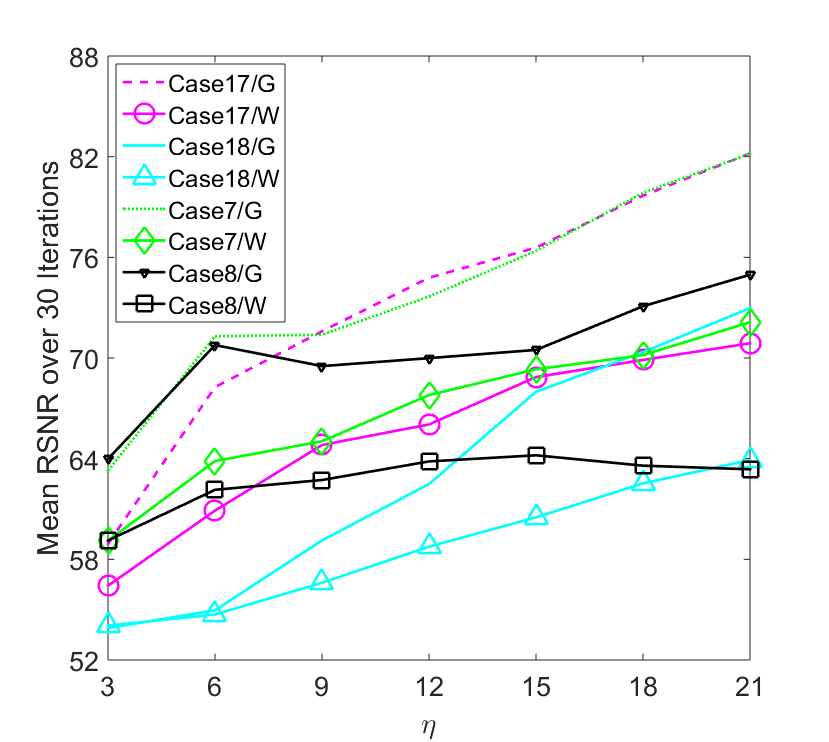}
        % Final Report >> finalfigs >> StephanieFigures >> Phantom_Data.png/fig/mat
        \caption{}
    \end{subfigure}
    \begin{subfigure}[h]{0.4\textwidth}
        \centering
        %\includegraphics[scale=0.33]{Brain_Data.png}
        % Final Report >> finalfigs >> StephanieFigures >> Brain_Data.png/fig/mat
				\includegraphics[scale=0.3]{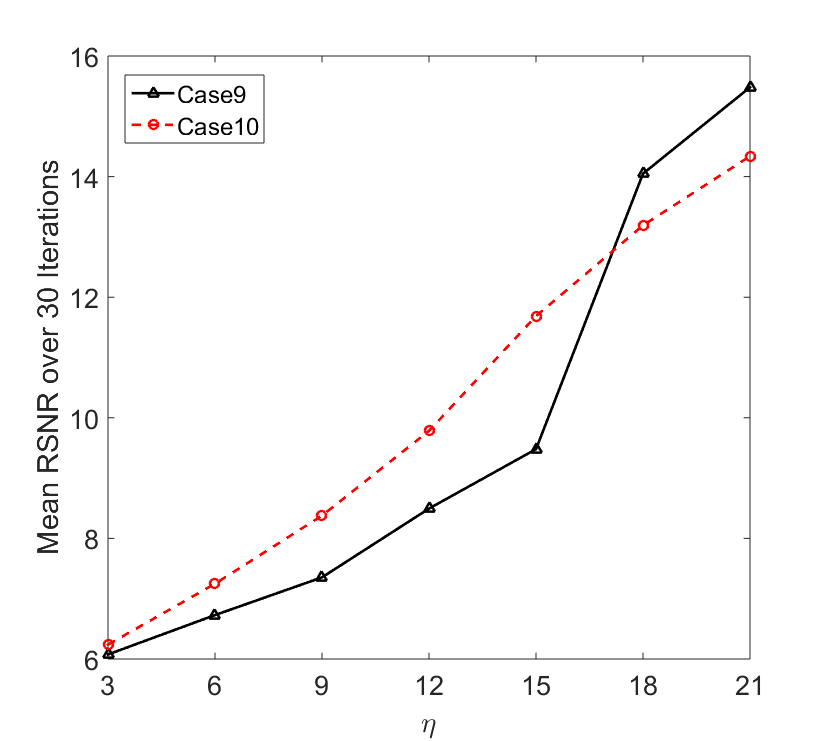}
        \caption{}
    \end{subfigure}
    \caption{(a) Reconstruction SNR of the recovered image at different compression levels for the phantom image ('G' and 'W' indicate the gradient and wavelet sparsifying transform used). (b) Reconstruction SNR of the recovered phantom image at different compression levels for cases 9 and 10 using L1-W (lower SNR).  $\eta$ is in \%. }
    \label{LACS_SNR}
\end{figure}

%%% Explain Case Simulations and Results %%%
We re-ran simulations for cases 7, 8, 17, and 18 on the phantom.  Over 30 trials, using the gradient on the phantom image resulted in much higher RSNR for every single case. Cases 7 and 17 both also outperformed cases 8 and 18 every time by about 13\%. These cases both used the variable density sampling method $f_{ VD}$ and LACS-MRI. Figure \ref{Phantom_Recovery} shows the reconstruction from Case 17. The recovered image using LACS-MRI with wavelet sparsity is more pixelated near the tumor while LACS-MRI with gradient sparsity produced an almost perfect reconstruction. 
Figure \ref{LACS_SNR} shows the great disparity between the two sparsifying methods. On average, the four phantom reconstructions using the gradient only needed to sample about 6\% of the data to achieve the same RSNR that the reconstructions using the wavelets did at 21\% sampling. 

%%% Figure: Reconstruction Using Case 17 for Phantom, Gradient vs. Wavelet %%%

%%% Figure: Reconstruction Using Case 10 for Phantom %%%
\begin{figure}[ht]
    \centering
    \begin{subfigure}[h]{0.3\textwidth}
        \centering
        \includegraphics[scale=0.35]{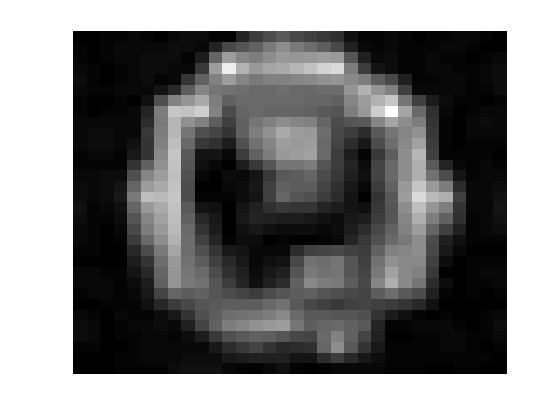}
        % Final Report >> finalfigs >> StephanieFigures >> Phantom_Case10_6.png/fig/mat
        \caption{}
    \end{subfigure}
    \begin{subfigure}[h]{0.3\textwidth}
        \centering
        \includegraphics[scale=0.35]{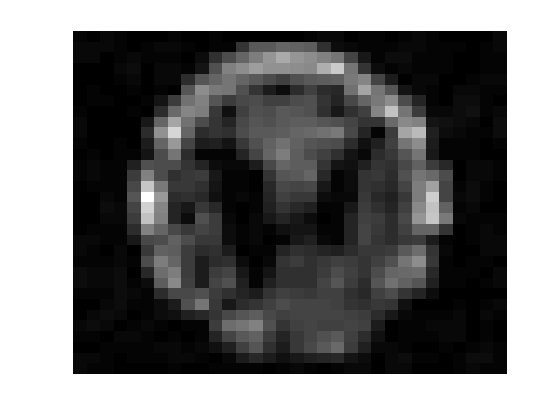}
        % Final Report >> finalfigs >> StephanieFigures >> Phantom_Case10_12.png/fig/mat
        \caption{}
    \end{subfigure}
    \begin{subfigure}[h]{0.3\textwidth}
        \centering
        \includegraphics[scale=0.35]{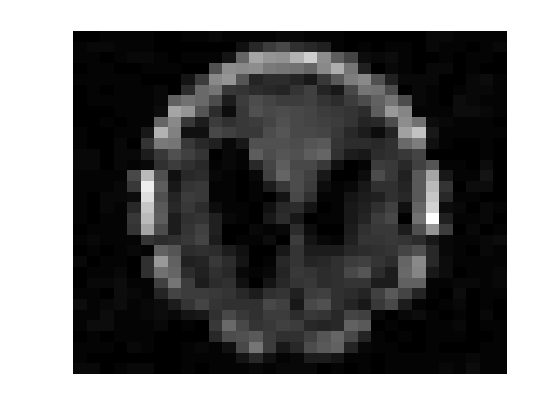}
        % Final Report >> finalfigs >> StephanieFigures >> Phantom_Case10_18.png/fig/mat
        \caption{}
    \end{subfigure}
    \caption{Case 10 using $\ell_{1}$-minimization with wavelet sparsification. Reconstruction with compression levels (a) 6 (b) 12 and (c) 18. The sampling methods used together were $f_{VDS}$ and $f_{R}$.}
    \label{Phantom_L1_Recovery}
\end{figure}

%%% Figure: Phantom and Brain LACS-MRI Reconstructions SNR Plot %%%

 We also compare the LACS-MRI gradient method to the Cases 9 and 10 which use $\ell_{1}$-minimization with wavelet sparsity. Clearly, the L1-W method recovered the phantom image very poorly, as shown in Figure \ref{Phantom_L1_Recovery} and Figure \ref{LACS_SNR} (b). The image is blurry even when 18\% of the available data was sampled. The RSNR of L1-W is worse than both the RSNR achieved by LACS-MRI with wavelet sparsity and  LACS-MRI with gradient sparsity.

We can conclude from these experiments that different sparsification methods can clearly have a drastic effect on the reconstruction of the follow-up image.  LACS-MRI with gradient sparsity is better for images with a sparser gradient than wavelet transform, such as the phantom image.  For images with high amounts of texture we expect the wavelet transform to work better.  
Therefore, if we know that an image will be more sparse in one domain over the other, we can appropriately choose the reconstruction method.  Such a determination of course may be non-trivial in some applications, and much research in compressed sensing and sparse approximation focuses on this judgement.  

\section{Grayscale Correction }

In a clinical setting, longitudinal MRI images may sometimes be taken on different MRI machines, on different days, with possibly different contrast settings, e.g.  ( T1 weighted, T2 weighted, Proton Density weighted) \cite{huang2014fast}.  This causes many practical issues in utilizing prior MRI scans as reference images; we tackle one of these problems, which is differing grayscale intensities.  We consider the problem where intensities of pixels of the same tissue are scaled very differently, and thus have differing amounts of contrast. This scaling factor can result in making the difference term, Term 2 in Equation \ref{optproblem}, be very large and thus the solution of the optimization problem would prioritize minimizing the difference term at the expense of fidelity to the actual measurements, resulting in low signal-to-noise ratio for the resultant image.  

\begin{figure*} [ht!]
    \centering
    \begin{subfigure}{0.5\textwidth}
        \centering
        \includegraphics[height=2.0in]{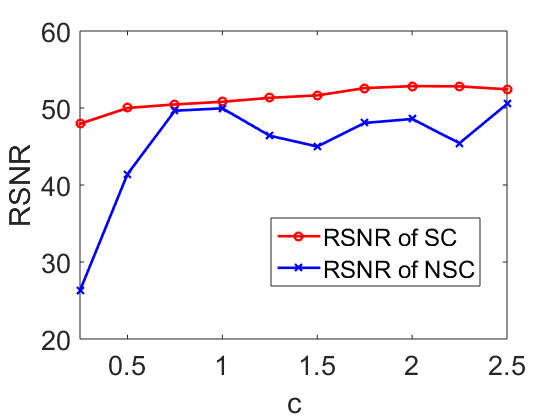}
        \caption{Brain Image}
    \end{subfigure}%
    ~ 
    \begin{subfigure}{0.5\textwidth}
        \centering
        \includegraphics[height=2.0in]{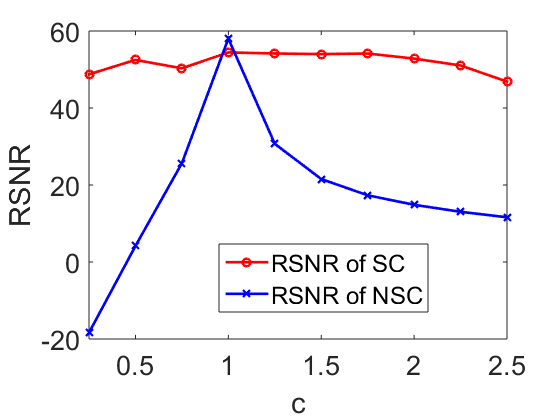}
        \caption{Shepp Logan Phantom}
    \end{subfigure}
    \caption{Reconstruction SNR vs $c$ for $c$= 0.25, 0.5, ... 2.5 with $\eta$=0.15 over 3 sampling iterations, each data point is the average of 50 trials.}
    \label{fig:c}
\end{figure*}

\begin{figure*}[ht!]
    \centering
    \begin{subfigure}{0.5\textwidth}
        \centering
        \includegraphics[height=2.0in]{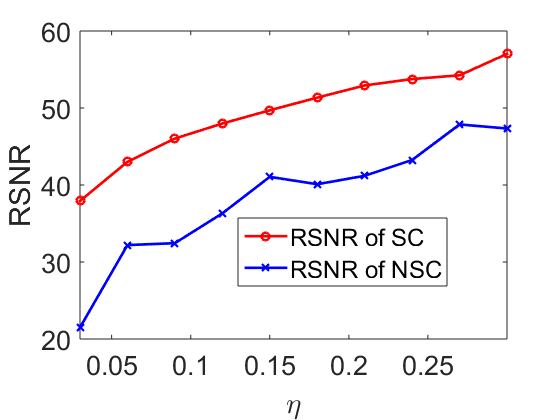}
        \caption{Brain Image}
    \end{subfigure}%
    ~ 
    \begin{subfigure}{0.5\textwidth}
        \centering
        \includegraphics[height=2.0in]{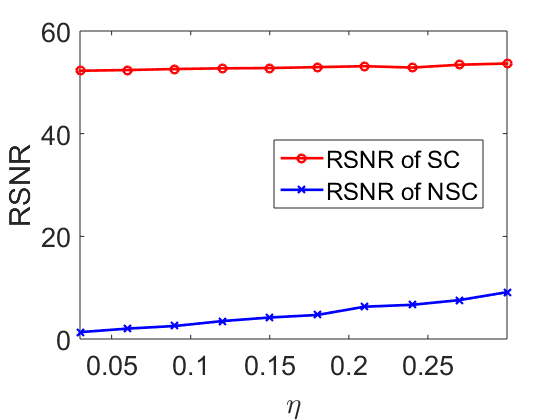}
        \caption{Shepp Logan Phantom}
    \end{subfigure}
    \caption{Reconstruction SNR vs $\eta$ for $\eta$= 0.03, 0.06, ... 0.30  over 3 sampling iterations with $c$=0.5, each data point is the average of 50 trials.}
    \label{fig:eta}
\end{figure*}

 The effect of the scale difference would be most pronounced in the argument of Term 2, but also impacts $\bW_1$ and $\bW_2$, since the values of these weighting terms  also depend on $\bx-\bx_0$. As shown in Figures \ref{fig:c}  and \ref{fig:eta}, the intensity shift causes significantly reduced RSNR. We have designed an algorithm and a modification of reference based MRI to compensate for the shift in grayscale intensity and ensure high RSNR. 

  We first  attempt a toy problem aimed at solving this issue, where we model the shift in grayscale intensity as linear, i.e. for some $c>0$, $\bX=c\bX_0$. If we know $c$ apriori, we can solve the optimization problem \eqref{optproblem} by modifying Term 2 in order to compensate for the intensity shift. Accordingly, Term 2 becomes $\lambda_2 \big \|W_2(\bX-c\bX_0) \big \|_1$.
This change corrects the inaccuracy of term 2 and allows the reconstruction to proceed based on the actual structure of the sampled image. We are now considering the correctly scaled difference between the reference and follow-up. The issue is how to estimate $c$.

\indent As  $\bX=c\bX_0$, we should have $c=\frac{\bX}{\bX_0}$. Since we only have access to the values of the Fourier coefficients obtained from the MRI, our proposed modified algorithm iteratively provides an estimate of $c$, based on the points sampled in each sampling round of LACS-MRI. The estimate of $c$ is then used to evaluate $c\bX_0$ in each iteration of the numerical method (rather than just $\bX_0$) to solve the original optimization problem \eqref{optproblem}. 
We define $\bY=\bF_{u}\bX$ and $\bY_{0}=\bF_{u}\bX_{0}$ to be the Fourier transforms of the follow-up and the reference image respectively, and describe our algorithmic modification by the following pseudocode. 

\begin{algorithm}[H]
\caption{Grayscale Correction for Reference Based MRI (GSC)}%\label{scale algorithm}
\begin{algorithmic}
\STATE \textbf{Input: }$\bY_{\samp}$ \text{, the sampling locations of the follow-up image}, $(\bY_{0})_{\samp}$\text{, the sampling locations of the reference image} $c$\text{ , the current grayscale estimate}, $i$, \text{number of sampling iterations}
\STATE $c^{'}= \frac{ \sum_{i,j }\bY_{i,j}}{\sum_{i,j }(\bY_0){i,j}}$
\STATE \textbf{Output:} $c=c^{'}\frac{1}{i} + c\frac{i-1}{i}$
\end{algorithmic}
\end{algorithm}

 The following plots clearly show the effectiveness of this scale correction method and the noisy results when scale is left uncorrected. Scale Correction (SC) refers to the reconstructions using the GSC algorithm, while No Scale Correction (NSC) is the standard unmodified LACS-MRI.

 \begin{figure*}[ht]    
 \centering 
 \begin{subfigure} {0.2\textwidth}
        \centering
        \includegraphics[height=1.75in,angle =90 ]{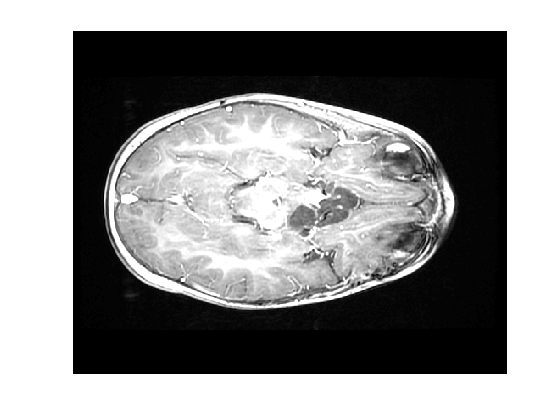}
        \caption{Ground Truth}
    \end{subfigure}
		 \begin{subfigure}{0.2\textwidth}
        \centering
        \includegraphics[height=1.75in,angle =90 ]{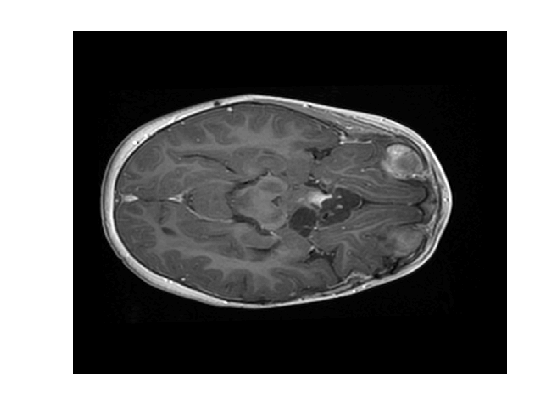}
        \caption{Reference}
    \end{subfigure}
\begin{subfigure}{0.2\textwidth}
        \centering
        \includegraphics[height=1.75in,angle =90 ]{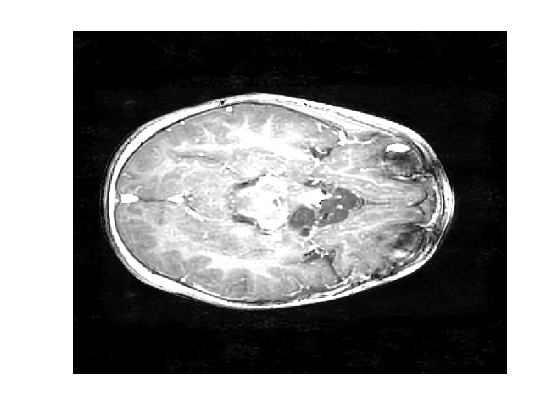}
        \caption{Scale Correction}
    \end{subfigure} 
  \begin{subfigure}{0.2\textwidth}
        \centering
        \includegraphics[height=1.75in,angle =90 ]{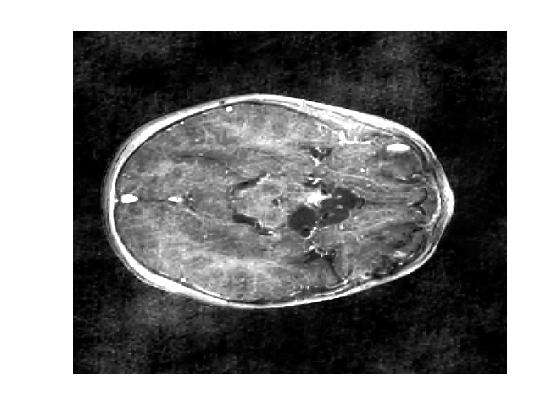}
        \caption{No Scale Correction}
    \end{subfigure}    
   
    \caption{(a) is the ground truth of the follow-up image, (b) is the reference image (with scaling parameter $c=2$), (c) and (d) are reconstructions with $\eta$=0.06 over 3 sampling iterations using LACS-MRI. }
    \label{recon}
\end{figure*}

Figure \ref{recon}  shows the effect of a 2-fold scale difference on LACS-MRI image reconstructions. There are extensive streaking artifacts and significant detail is lost compared to the ground truth and the reconstruction using the modified method. Clearly, the Grayscale Compensation for Reference Based MRI Algorithm (GSC) can reduce these artifacts and help improve the visibility of structure.

 Figure \ref{fig:c} shows the relationship of RSNR and $c$. Figure \ref{fig:c} is validated by the fact that when $c=1$, the RSNR of both the Brain image and Shepp Logan Phantom converge to the same value. There is some residual error due to noise in the estimation of c. 

 Figure \ref{fig:eta} shows the relationship between RSNR and $\eta$, where $\eta$ is the total amount of the source image sampled. As to be expected, in both cases RSNR increases with increasing $\eta$, or more samples will yield  better reconstructions. 
\FloatBarrier

 These results support the validity of our solution to this toy problem and let us hope that similar but less naive techniques can be applied to more complex issues, i.e. variable contrast and scaling localized to sub-regions of brain scans. As in the linear toy problem, subregions exhibiting this type of behavior can result in lowered RSNR. The solution to this larger problem would greatly enhance the value of LACS-MRI in clinical situations and for greater robustness to variation in MRI acquisition parameters.

\section{Conclusion}

We have built upon previous work done by Weizman et. al. \cite{weizman2015compressed} by changing the sampling method, applying gradient as a sparsifier, and compensating for differences in grayscale level between reference and follow-up images. In addition, we investigated the combinations of different adaptive sampling functions, variable sampling functions, and reconstruction algorithms in an attempt to find an optimal image reconstruction method. We found that the accuracy of resultant reconstructions is dependent on the properties of the image itself, in a certain basis, and that taking advantage of this knowledge can yield high RSNR given minimal samples. 
	\\ \indent We hope to pursue future research in more complex non-linear grayscale compensation issues where only a portion of the image is scaled differently.  We also hope to find theoretical bounds on these reference based methods applied to different classes of images.

\bibliographystyle{acm}
 %\clearpage
\bibliography{finalbib}

\begin{thebibliography}{10}

\bibitem{berman1988matrix}
{\sc Berman, K., Halpern, H., Kaftal, V., and Weiss, G.}
\newblock Matrix norm inequalities and the relative dixmier property.
\newblock {\em Integral Equations and Operator Theory 11}, 1 (1988), 28--48.

\bibitem{candes2005error}
{\sc Candes, E., Rudelson, M., Tao, T., and Vershynin, R.}
\newblock Error correction via linear programming.
\newblock In {\em 46th Annual IEEE Symposium on Foundations of Computer Science
  (FOCS'05)\/} (2005), IEEE, pp.~668--681.

\bibitem{RefWorks:309}
{\sc Cand\`es, E.~J.}
\newblock Compressive sampling.
\newblock In {\em Proc. Int. Congress of Mathematics\/} (Madrid, Spain, 2006),
  vol.~3, pp.~1433--1452.

\bibitem{emmanuel2004robust}
{\sc Cand{\`e}s, E.~J., Romberg, J., and Tao, T.}
\newblock Robust uncertainty principles: Exact signal reconstruction from
  highly incomplete frequency information.
\newblock {\em IEEE Trans. Inform. Theory 52}, 2 (2006), 489--509.

\bibitem{candes2006stable}
{\sc Candes, E.~J., Romberg, J.~K., and Tao, T.}
\newblock Stable signal recovery from incomplete and inaccurate measurements.
\newblock {\em Communications on pure and applied mathematics 59}, 8 (2006),
  1207--1223.

\bibitem{Daube_Ten}
{\sc Daubechies, I.}
\newblock {\em Ten lectures on wavelets}.
\newblock SIAM, Philadelphia, PA, 1992.

\bibitem{davenport2015constrained}
{\sc Davenport, M.~A., Massimino, A.~K., Needell, D., and Woolf, T.}
\newblock Constrained adaptive sensing.
\newblock {\em arXiv preprint arXiv:1506.05889\/} (2015).

\bibitem{RefWorks:70}
{\sc Donoho, D.~L.}
\newblock Compressed sensing.
\newblock {\em IEEE T. Inform. Theory 52}, 4 (Apr. 2006), 1289--1306.

\bibitem{hollingworth2000diagnostic}
{\sc Hollingworth, W., Todd, C.~J., Bell, M.~I., Arafat, Q., Girling, S.,
  Karia, K.~R., and Dixon, A.~K.}
\newblock The diagnostic and therapeutic impact of mri: an observational
  multi-centre study.
\newblock {\em Clinical radiology 55}, 11 (2000), 825--831.

\bibitem{huang2014fast}
{\sc Huang, J., Chen, C., and Axel, L.}
\newblock Fast multi-contrast mri reconstruction.
\newblock {\em Magnetic resonance imaging 32}, 10 (2014), 1344--1352.

\bibitem{kirschen1989use}
{\sc Kirschen, D.~G.}
\newblock Use in environment of an mri scanner, Feb.~14 1989.
\newblock US Patent 4,804,261.

\bibitem{krahmer2014stable}
{\sc Krahmer, F., and Ward, R.}
\newblock Stable and robust sampling strategies for compressive imaging.
\newblock {\em Image Processing, IEEE Transactions on 23}, 2 (2014), 612--622.

\bibitem{lustig2007sparse}
{\sc Lustig, M., Donoho, D., and Pauly, J.~M.}
\newblock Sparse mri: The application of compressed sensing for rapid mr
  imaging.
\newblock {\em Magnetic resonance in medicine 58}, 6 (2007), 1182--1195.

\bibitem{lustig2010spirit}
{\sc Lustig, M., and Pauly, J.~M.}
\newblock Spirit: Iterative self-consistent parallel imaging reconstruction
  from arbitrary k-space.
\newblock {\em Magnetic Resonance in Medicine 64}, 2 (2010), 457--471.

\bibitem{muthukrishnan2005data}
{\sc Muthukrishnan, S.}
\newblock {\em Data streams: Algorithms and applications}.
\newblock Now Publishers Inc, 2005.

\bibitem{RefWorks:7}
{\sc Needell, D., and Ward, R.}
\newblock Stable image reconstruction using total variation minimization.
\newblock {\em SIAM Journal on Imaging Sciences 6}, 2 (2013), 1035--1058.

\bibitem{RauhuSV_Compressed}
{\sc Rauhut, H., Schnass, K., and Vandergheynst, P.}
\newblock Compressed sensing and redundant dictionaries.
\newblock {\em IEEE Trans. Inform. Theory 54}, 5 (2008), 2210--2219.

\bibitem{RefWorks:285}
{\sc Rudelson, M., and Vershynin, R.}
\newblock On sparse reconstruction from fourier and gaussian measurements.
\newblock {\em Comm. Pure Appl. Math. 61\/} (2008), 1025--1045.

\bibitem{weizman2014application}
{\sc Weizman, L., Eldar, Y.~C., and Bashat, D.~B.}
\newblock The application of compressed sensing for longitudinal mri.
\newblock {\em arXiv preprint arXiv:1407.2602\/} (2014).

\bibitem{weizman2015compressed}
{\sc Weizman, L., Eldar, Y.~C., and Bashat, D.~B.}
\newblock Compressed sensing for longitudinal mri: An adaptive-weighted
  approach.
\newblock {\em Medical physics 42}, 9 (2015), 5195--5208.

\end{thebibliography}

\end{document}